\pdfoutput=1
\documentclass{llncs}

\usepackage[FIGBOTCAP,TABBOTCAP,center,nooneline]{subfigure}
\usepackage[dvips]{rotating}
\usepackage{RSalgorithmic}
\usepackage{algorithm}
\usepackage{amscd}
\usepackage{amsfonts}
\usepackage{amsmath}
\usepackage{amssymb}
\usepackage{booktabs}
\usepackage{color}
\usepackage{comment}
\usepackage{datetime}
\usepackage{epsfig}
\usepackage{graphicx}
\usepackage{latexsym}
\usepackage{marginnote}
\usepackage{multirow}
\usepackage{paralist}
\usepackage{placeins}
\usepackage{rotate} 
\usepackage[FIGBOTCAP,TABBOTCAP,center,nooneline]{subfigure}
\usepackage{tikz}
\usepackage[disable]{todonotes}
\usepackage{times} 
\usepackage{url}
\usepackage{varwidth}
\usepackage{verbatim} 
\usepackage{wrapfig}
\usepackage{xcolor,colortbl}
\usepackage{xspace}

\definecolor {snow}                {rgb}{1.00,0.98,0.98}
\definecolor {ghostwhite}          {rgb}{0.97,0.97,1.00}
\definecolor {whitesmoke}          {rgb}{0.96,0.96,0.96}
\definecolor {gainsboro}           {rgb}{0.86,0.86,0.86}
\definecolor {floralwhite}         {rgb}{1.00,0.98,0.94}
\definecolor {oldlace}             {rgb}{0.99,0.96,0.90}
\definecolor {linen}               {rgb}{0.98,0.94,0.90}
\definecolor {antiquewhite}        {rgb}{0.98,0.92,0.84}
\definecolor {papayawhip}          {rgb}{1.00,0.94,0.84}
\definecolor {blanchedalmond}      {rgb}{1.00,0.92,0.80}
\definecolor {bisque}              {rgb}{1.00,0.89,0.77}
\definecolor {peachpuff}           {rgb}{1.00,0.85,0.73}
\definecolor {navajowhite}         {rgb}{1.00,0.87,0.68}
\definecolor {moccasin}            {rgb}{1.00,0.89,0.71}
\definecolor {cornsilk}            {rgb}{1.00,0.97,0.86}
\definecolor {ivory}               {rgb}{1.00,1.00,0.94}
\definecolor {lemonchiffon}        {rgb}{1.00,0.98,0.80}
\definecolor {seashell}            {rgb}{1.00,0.96,0.93}
\definecolor {honeydew}            {rgb}{0.94,1.00,0.94}
\definecolor {mintcream}           {rgb}{0.96,1.00,0.98}
\definecolor {azure}               {rgb}{0.94,1.00,1.00}
\definecolor {aliceblue}           {rgb}{0.94,0.97,1.00}
\definecolor {lavender}            {rgb}{0.90,0.90,0.98}
\definecolor {lavenderblush}       {rgb}{1.00,0.94,0.96}
\definecolor {mistyrose}           {rgb}{1.00,0.89,0.88}
\definecolor {white}               {rgb}{1.00,1.00,1.00}
\definecolor {black}               {rgb}{0.00,0.00,0.00}
\definecolor {darkslategray}       {rgb}{0.18,0.31,0.31}
\definecolor {dimgray}             {rgb}{0.41,0.41,0.41}
\definecolor {slategray}           {rgb}{0.44,0.50,0.56}
\definecolor {lightslategray}      {rgb}{0.47,0.53,0.60}
\definecolor {gray}                {rgb}{0.75,0.75,0.75}
\definecolor {lightgrey}           {rgb}{0.83,0.83,0.83}
\definecolor {midnightblue}        {rgb}{0.10,0.10,0.44}
\definecolor {navy}                {rgb}{0.00,0.00,0.50}
\definecolor {cornflowerblue}      {rgb}{0.39,0.58,0.93}
\definecolor {darkslateblue}       {rgb}{0.28,0.24,0.55}
\definecolor {slateblue}           {rgb}{0.42,0.35,0.80}
\definecolor {mediumslateblue}     {rgb}{0.48,0.41,0.93}
\definecolor {lightslateblue}      {rgb}{0.52,0.44,1.00}
\definecolor {mediumblue}          {rgb}{0.00,0.00,0.80}
\definecolor {royalblue}           {rgb}{0.25,0.41,0.88}
\definecolor {blue}                {rgb}{0.00,0.00,1.00}
\definecolor {dodgerblue}          {rgb}{0.12,0.56,1.00}
\definecolor {deepskyblue}         {rgb}{0.00,0.75,1.00}
\definecolor {skyblue}             {rgb}{0.53,0.81,0.92}
\definecolor {lightskyblue}        {rgb}{0.53,0.81,0.98}
\definecolor {steelblue}           {rgb}{0.27,0.51,0.71}
\definecolor {lightsteelblue}      {rgb}{0.69,0.77,0.87}
\definecolor {lightblue}           {rgb}{0.68,0.85,0.90}
\definecolor {powderblue}          {rgb}{0.69,0.88,0.90}
\definecolor {paleturquoise}       {rgb}{0.69,0.93,0.93}
\definecolor {darkturquoise}       {rgb}{0.00,0.81,0.82}
\definecolor {mediumturquoise}     {rgb}{0.28,0.82,0.80}
\definecolor {turquoise}           {rgb}{0.25,0.88,0.82}
\definecolor {cyan}                {rgb}{0.00,1.00,1.00}
\definecolor {lightcyan}           {rgb}{0.88,1.00,1.00}
\definecolor {cadetblue}           {rgb}{0.37,0.62,0.63}
\definecolor {mediumaquamarine}    {rgb}{0.40,0.80,0.67}
\definecolor {aquamarine}          {rgb}{0.50,1.00,0.83}
\definecolor {darkgreen}           {rgb}{0.00,0.39,0.00}
\definecolor {darkolivegreen}      {rgb}{0.33,0.42,0.18}
\definecolor {darkseagreen}        {rgb}{0.56,0.74,0.56}
\definecolor {seagreen}            {rgb}{0.18,0.55,0.34}
\definecolor {mediumseagreen}      {rgb}{0.24,0.70,0.44}
\definecolor {lightseagreen}       {rgb}{0.13,0.70,0.67}
\definecolor {palegreen}           {rgb}{0.60,0.98,0.60}
\definecolor {springgreen}         {rgb}{0.00,1.00,0.50}
\definecolor {lawngreen}           {rgb}{0.49,0.99,0.00}
\definecolor {green}               {rgb}{0.00,1.00,0.00}
\definecolor {chartreuse}          {rgb}{0.50,1.00,0.00}
\definecolor {mediumspringgreen}   {rgb}{0.00,0.98,0.60}
\definecolor {greenyellow}         {rgb}{0.68,1.00,0.18}
\definecolor {limegreen}           {rgb}{0.20,0.80,0.20}
\definecolor {yellowgreen}         {rgb}{0.60,0.80,0.20}
\definecolor {forestgreen}         {rgb}{0.13,0.55,0.13}
\definecolor {olivedrab}           {rgb}{0.42,0.56,0.14}
\definecolor {darkkhaki}           {rgb}{0.74,0.72,0.42}
\definecolor {khaki}               {rgb}{0.94,0.90,0.55}
\definecolor {palegoldenrod}       {rgb}{0.93,0.91,0.67}
\definecolor {lightgoldenrodyellow} {rgb}{0.98,0.98,0.82}
\definecolor {lightyellow}         {rgb}{1.00,1.00,0.88}
\definecolor {yellow}              {rgb}{1.00,1.00,0.00}
\definecolor {gold}                {rgb}{1.00,0.84,0.00}
\definecolor {lightgoldenrod}      {rgb}{0.93,0.87,0.51}
\definecolor {goldenrod}           {rgb}{0.85,0.65,0.13}
\definecolor {darkgoldenrod}       {rgb}{0.72,0.53,0.04}
\definecolor {rosybrown}           {rgb}{0.74,0.56,0.56}
\definecolor {indianred}           {rgb}{0.80,0.36,0.36}
\definecolor {saddlebrown}         {rgb}{0.55,0.27,0.07}
\definecolor {sienna}              {rgb}{0.63,0.32,0.18}
\definecolor {peru}                {rgb}{0.80,0.52,0.25}
\definecolor {burlywood}           {rgb}{0.87,0.72,0.53}
\definecolor {beige}               {rgb}{0.96,0.96,0.86}
\definecolor {wheat}               {rgb}{0.96,0.87,0.70}
\definecolor {sandybrown}          {rgb}{0.96,0.64,0.38}
\definecolor {tan}                 {rgb}{0.82,0.71,0.55}
\definecolor {chocolate}           {rgb}{0.82,0.41,0.12}
\definecolor {firebrick}           {rgb}{0.70,0.13,0.13}
\definecolor {brown}               {rgb}{0.65,0.16,0.16}
\definecolor {darksalmon}          {rgb}{0.91,0.59,0.48}
\definecolor {salmon}              {rgb}{0.98,0.50,0.45}
\definecolor {lightsalmon}         {rgb}{1.00,0.63,0.48}
\definecolor {orange}              {rgb}{1.00,0.65,0.00}
\definecolor {darkorange}          {rgb}{1.00,0.55,0.00}
\definecolor {coral}               {rgb}{1.00,0.50,0.31}
\definecolor {lightcoral}          {rgb}{0.94,0.50,0.50}
\definecolor {tomato}              {rgb}{1.00,0.39,0.28}
\definecolor {orangered}           {rgb}{1.00,0.27,0.00}
\definecolor {red}                 {rgb}{1.00,0.00,0.00}
\definecolor {hotpink}             {rgb}{1.00,0.41,0.71}
\definecolor {deeppink}            {rgb}{1.00,0.08,0.58}
\definecolor {pink}                {rgb}{1.00,0.75,0.80}
\definecolor {lightpink}           {rgb}{1.00,0.71,0.76}
\definecolor {palevioletred}       {rgb}{0.86,0.44,0.58}
\definecolor {maroon}              {rgb}{0.69,0.19,0.38}
\definecolor {mediumvioletred}     {rgb}{0.78,0.08,0.52}
\definecolor {violetred}           {rgb}{0.82,0.13,0.56}
\definecolor {magenta}             {rgb}{1.00,0.00,1.00}
\definecolor {violet}              {rgb}{0.93,0.51,0.93}
\definecolor {plum}                {rgb}{0.87,0.63,0.87}
\definecolor {orchid}              {rgb}{0.85,0.44,0.84}
\definecolor {mediumorchid}        {rgb}{0.73,0.33,0.83}
\definecolor {darkorchid}          {rgb}{0.60,0.20,0.80}
\definecolor {darkviolet}          {rgb}{0.58,0.00,0.83}
\definecolor {blueviolet}          {rgb}{0.54,0.17,0.89}
\definecolor {purple}              {rgb}{0.63,0.13,0.94}
\definecolor {mediumpurple}        {rgb}{0.58,0.44,0.86}
\definecolor {thistle}             {rgb}{0.85,0.75,0.85}
\definecolor {snow2}               {rgb}{0.93,0.91,0.91}
\definecolor {snow3}               {rgb}{0.80,0.79,0.79}
\definecolor {snow4}               {rgb}{0.55,0.54,0.54}
\definecolor {seashell2}           {rgb}{0.93,0.90,0.87}
\definecolor {seashell3}           {rgb}{0.80,0.77,0.75}
\definecolor {seashell4}           {rgb}{0.55,0.53,0.51}
\definecolor {antiquewhite1}       {rgb}{1.00,0.94,0.86}
\definecolor {antiquewhite2}       {rgb}{0.93,0.87,0.80}
\definecolor {antiquewhite3}       {rgb}{0.80,0.75,0.69}
\definecolor {antiquewhite4}       {rgb}{0.55,0.51,0.47}
\definecolor {bisque2}             {rgb}{0.93,0.84,0.72}
\definecolor {bisque3}             {rgb}{0.80,0.72,0.62}
\definecolor {bisque4}             {rgb}{0.55,0.49,0.42}
\definecolor {peachpuff2}          {rgb}{0.93,0.80,0.68}
\definecolor {peachpuff3}          {rgb}{0.80,0.69,0.58}
\definecolor {peachpuff4}          {rgb}{0.55,0.47,0.40}
\definecolor {navajowhite2}        {rgb}{0.93,0.81,0.63}
\definecolor {navajowhite3}        {rgb}{0.80,0.70,0.55}
\definecolor {navajowhite4}        {rgb}{0.55,0.47,0.37}
\definecolor {lemonchiffon2}       {rgb}{0.93,0.91,0.75}
\definecolor {lemonchiffon3}       {rgb}{0.80,0.79,0.65}
\definecolor {lemonchiffon4}       {rgb}{0.55,0.54,0.44}
\definecolor {cornsilk2}           {rgb}{0.93,0.91,0.80}
\definecolor {cornsilk3}           {rgb}{0.80,0.78,0.69}
\definecolor {cornsilk4}           {rgb}{0.55,0.53,0.47}
\definecolor {ivory2}              {rgb}{0.93,0.93,0.88}
\definecolor {ivory3}              {rgb}{0.80,0.80,0.76}
\definecolor {ivory4}              {rgb}{0.55,0.55,0.51}
\definecolor {honeydew2}           {rgb}{0.88,0.93,0.88}
\definecolor {honeydew3}           {rgb}{0.76,0.80,0.76}
\definecolor {honeydew4}           {rgb}{0.51,0.55,0.51}
\definecolor {lavenderblush2}      {rgb}{0.93,0.88,0.90}
\definecolor {lavenderblush3}      {rgb}{0.80,0.76,0.77}
\definecolor {lavenderblush4}      {rgb}{0.55,0.51,0.53}
\definecolor {mistyrose2}          {rgb}{0.93,0.84,0.82}
\definecolor {mistyrose3}          {rgb}{0.80,0.72,0.71}
\definecolor {mistyrose4}          {rgb}{0.55,0.49,0.48}
\definecolor {azure2}              {rgb}{0.88,0.93,0.93}
\definecolor {azure3}              {rgb}{0.76,0.80,0.80}
\definecolor {azure4}              {rgb}{0.51,0.55,0.55}
\definecolor {slateblue1}          {rgb}{0.51,0.44,1.00}
\definecolor {slateblue2}          {rgb}{0.48,0.40,0.93}
\definecolor {slateblue3}          {rgb}{0.41,0.35,0.80}
\definecolor {slateblue4}          {rgb}{0.28,0.24,0.55}
\definecolor {royalblue1}          {rgb}{0.28,0.46,1.00}
\definecolor {royalblue2}          {rgb}{0.26,0.43,0.93}
\definecolor {royalblue3}          {rgb}{0.23,0.37,0.80}
\definecolor {royalblue4}          {rgb}{0.15,0.25,0.55}
\definecolor {blue2}               {rgb}{0.00,0.00,0.93}
\definecolor {blue4}               {rgb}{0.00,0.00,0.55}
\definecolor {dodgerblue2}         {rgb}{0.11,0.53,0.93}
\definecolor {dodgerblue3}         {rgb}{0.09,0.45,0.80}
\definecolor {dodgerblue4}         {rgb}{0.06,0.31,0.55}
\definecolor {steelblue1}          {rgb}{0.39,0.72,1.00}
\definecolor {steelblue2}          {rgb}{0.36,0.67,0.93}
\definecolor {steelblue3}          {rgb}{0.31,0.58,0.80}
\definecolor {steelblue4}          {rgb}{0.21,0.39,0.55}
\definecolor {deepskyblue2}        {rgb}{0.00,0.70,0.93}
\definecolor {deepskyblue3}        {rgb}{0.00,0.60,0.80}
\definecolor {deepskyblue4}        {rgb}{0.00,0.41,0.55}
\definecolor {skyblue1}            {rgb}{0.53,0.81,1.00}
\definecolor {skyblue2}            {rgb}{0.49,0.75,0.93}
\definecolor {skyblue3}            {rgb}{0.42,0.65,0.80}
\definecolor {skyblue4}            {rgb}{0.29,0.44,0.55}
\definecolor {lightskyblue1}       {rgb}{0.69,0.89,1.00}
\definecolor {lightskyblue2}       {rgb}{0.64,0.83,0.93}
\definecolor {lightskyblue3}       {rgb}{0.55,0.71,0.80}
\definecolor {lightskyblue4}       {rgb}{0.38,0.48,0.55}
\definecolor {slategray1}          {rgb}{0.78,0.89,1.00}
\definecolor {slategray2}          {rgb}{0.73,0.83,0.93}
\definecolor {slategray3}          {rgb}{0.62,0.71,0.80}
\definecolor {slategray4}          {rgb}{0.42,0.48,0.55}
\definecolor {lightsteelblue1}     {rgb}{0.79,0.88,1.00}
\definecolor {lightsteelblue2}     {rgb}{0.74,0.82,0.93}
\definecolor {lightsteelblue3}     {rgb}{0.64,0.71,0.80}
\definecolor {lightsteelblue4}     {rgb}{0.43,0.48,0.55}
\definecolor {lightblue1}          {rgb}{0.75,0.94,1.00}
\definecolor {lightblue2}          {rgb}{0.70,0.87,0.93}
\definecolor {lightblue3}          {rgb}{0.60,0.75,0.80}
\definecolor {lightblue4}          {rgb}{0.41,0.51,0.55}
\definecolor {lightcyan2}          {rgb}{0.82,0.93,0.93}
\definecolor {lightcyan3}          {rgb}{0.71,0.80,0.80}
\definecolor {lightcyan4}          {rgb}{0.48,0.55,0.55}
\definecolor {paleturquoise1}      {rgb}{0.73,1.00,1.00}
\definecolor {paleturquoise2}      {rgb}{0.68,0.93,0.93}
\definecolor {paleturquoise3}      {rgb}{0.59,0.80,0.80}
\definecolor {paleturquoise4}      {rgb}{0.40,0.55,0.55}
\definecolor {cadetblue1}          {rgb}{0.60,0.96,1.00}
\definecolor {cadetblue2}          {rgb}{0.56,0.90,0.93}
\definecolor {cadetblue3}          {rgb}{0.48,0.77,0.80}
\definecolor {cadetblue4}          {rgb}{0.33,0.53,0.55}
\definecolor {turquoise1}          {rgb}{0.00,0.96,1.00}
\definecolor {turquoise2}          {rgb}{0.00,0.90,0.93}
\definecolor {turquoise3}          {rgb}{0.00,0.77,0.80}
\definecolor {turquoise4}          {rgb}{0.00,0.53,0.55}
\definecolor {cyan2}               {rgb}{0.00,0.93,0.93}
\definecolor {cyan3}               {rgb}{0.00,0.80,0.80}
\definecolor {cyan4}               {rgb}{0.00,0.55,0.55}
\definecolor {darkslategray1}      {rgb}{0.59,1.00,1.00}
\definecolor {darkslategray2}      {rgb}{0.55,0.93,0.93}
\definecolor {darkslategray3}      {rgb}{0.47,0.80,0.80}
\definecolor {darkslategray4}      {rgb}{0.32,0.55,0.55}
\definecolor {aquamarine2}         {rgb}{0.46,0.93,0.78}
\definecolor {aquamarine4}         {rgb}{0.27,0.55,0.45}
\definecolor {darkseagreen1}       {rgb}{0.76,1.00,0.76}
\definecolor {darkseagreen2}       {rgb}{0.71,0.93,0.71}
\definecolor {darkseagreen3}       {rgb}{0.61,0.80,0.61}
\definecolor {darkseagreen4}       {rgb}{0.41,0.55,0.41}
\definecolor {seagreen1}           {rgb}{0.33,1.00,0.62}
\definecolor {seagreen2}           {rgb}{0.31,0.93,0.58}
\definecolor {seagreen3}           {rgb}{0.26,0.80,0.50}
\definecolor {palegreen1}          {rgb}{0.60,1.00,0.60}
\definecolor {palegreen2}          {rgb}{0.56,0.93,0.56}
\definecolor {palegreen3}          {rgb}{0.49,0.80,0.49}
\definecolor {palegreen4}          {rgb}{0.33,0.55,0.33}
\definecolor {springgreen2}        {rgb}{0.00,0.93,0.46}
\definecolor {springgreen3}        {rgb}{0.00,0.80,0.40}
\definecolor {springgreen4}        {rgb}{0.00,0.55,0.27}
\definecolor {green2}              {rgb}{0.00,0.93,0.00}
\definecolor {green3}              {rgb}{0.00,0.80,0.00}
\definecolor {green4}              {rgb}{0.00,0.55,0.00}
\definecolor {chartreuse2}         {rgb}{0.46,0.93,0.00}
\definecolor {chartreuse3}         {rgb}{0.40,0.80,0.00}
\definecolor {chartreuse4}         {rgb}{0.27,0.55,0.00}
\definecolor {olivedrab1}          {rgb}{0.75,1.00,0.24}
\definecolor {olivedrab2}          {rgb}{0.70,0.93,0.23}
\definecolor {olivedrab4}          {rgb}{0.41,0.55,0.13}
\definecolor {darkolivegreen1}     {rgb}{0.79,1.00,0.44}
\definecolor {darkolivegreen2}     {rgb}{0.74,0.93,0.41}
\definecolor {darkolivegreen3}     {rgb}{0.64,0.80,0.35}
\definecolor {darkolivegreen4}     {rgb}{0.43,0.55,0.24}
\definecolor {khaki1}              {rgb}{1.00,0.96,0.56}
\definecolor {khaki2}              {rgb}{0.93,0.90,0.52}
\definecolor {khaki3}              {rgb}{0.80,0.78,0.45}
\definecolor {khaki4}              {rgb}{0.55,0.53,0.31}
\definecolor {lightgoldenrod1}     {rgb}{1.00,0.93,0.55}
\definecolor {lightgoldenrod2}     {rgb}{0.93,0.86,0.51}
\definecolor {lightgoldenrod3}     {rgb}{0.80,0.75,0.44}
\definecolor {lightgoldenrod4}     {rgb}{0.55,0.51,0.30}
\definecolor {lightyellow2}        {rgb}{0.93,0.93,0.82}
\definecolor {lightyellow3}        {rgb}{0.80,0.80,0.71}
\definecolor {lightyellow4}        {rgb}{0.55,0.55,0.48}
\definecolor {yellow2}             {rgb}{0.93,0.93,0.00}
\definecolor {yellow3}             {rgb}{0.80,0.80,0.00}
\definecolor {yellow4}             {rgb}{0.55,0.55,0.00}
\definecolor {gold2}               {rgb}{0.93,0.79,0.00}
\definecolor {gold3}               {rgb}{0.80,0.68,0.00}
\definecolor {gold4}               {rgb}{0.55,0.46,0.00}
\definecolor {goldenrod1}          {rgb}{1.00,0.76,0.15}
\definecolor {goldenrod2}          {rgb}{0.93,0.71,0.13}
\definecolor {goldenrod3}          {rgb}{0.80,0.61,0.11}
\definecolor {goldenrod4}          {rgb}{0.55,0.41,0.08}
\definecolor {darkgoldenrod1}      {rgb}{1.00,0.73,0.06}
\definecolor {darkgoldenrod2}      {rgb}{0.93,0.68,0.05}
\definecolor {darkgoldenrod3}      {rgb}{0.80,0.58,0.05}
\definecolor {darkgoldenrod4}      {rgb}{0.55,0.40,0.03}
\definecolor {rosybrown1}          {rgb}{1.00,0.76,0.76}
\definecolor {rosybrown2}          {rgb}{0.93,0.71,0.71}
\definecolor {rosybrown3}          {rgb}{0.80,0.61,0.61}
\definecolor {rosybrown4}          {rgb}{0.55,0.41,0.41}
\definecolor {indianred1}          {rgb}{1.00,0.42,0.42}
\definecolor {indianred2}          {rgb}{0.93,0.39,0.39}
\definecolor {indianred3}          {rgb}{0.80,0.33,0.33}
\definecolor {indianred4}          {rgb}{0.55,0.23,0.23}
\definecolor {sienna1}             {rgb}{1.00,0.51,0.28}
\definecolor {sienna2}             {rgb}{0.93,0.47,0.26}
\definecolor {sienna3}             {rgb}{0.80,0.41,0.22}
\definecolor {sienna4}             {rgb}{0.55,0.28,0.15}
\definecolor {burlywood1}          {rgb}{1.00,0.83,0.61}
\definecolor {burlywood2}          {rgb}{0.93,0.77,0.57}
\definecolor {burlywood3}          {rgb}{0.80,0.67,0.49}
\definecolor {burlywood4}          {rgb}{0.55,0.45,0.33}
\definecolor {wheat1}              {rgb}{1.00,0.91,0.73}
\definecolor {wheat2}              {rgb}{0.93,0.85,0.68}
\definecolor {wheat3}              {rgb}{0.80,0.73,0.59}
\definecolor {wheat4}              {rgb}{0.55,0.49,0.40}
\definecolor {tan1}                {rgb}{1.00,0.65,0.31}
\definecolor {tan2}                {rgb}{0.93,0.60,0.29}
\definecolor {tan4}                {rgb}{0.55,0.35,0.17}
\definecolor {chocolate1}          {rgb}{1.00,0.50,0.14}
\definecolor {chocolate2}          {rgb}{0.93,0.46,0.13}
\definecolor {chocolate3}          {rgb}{0.80,0.40,0.11}
\definecolor {firebrick1}          {rgb}{1.00,0.19,0.19}
\definecolor {firebrick2}          {rgb}{0.93,0.17,0.17}
\definecolor {firebrick3}          {rgb}{0.80,0.15,0.15}
\definecolor {firebrick4}          {rgb}{0.55,0.10,0.10}
\definecolor {brown1}              {rgb}{1.00,0.25,0.25}
\definecolor {brown2}              {rgb}{0.93,0.23,0.23}
\definecolor {brown3}              {rgb}{0.80,0.20,0.20}
\definecolor {brown4}              {rgb}{0.55,0.14,0.14}
\definecolor {salmon1}             {rgb}{1.00,0.55,0.41}
\definecolor {salmon2}             {rgb}{0.93,0.51,0.38}
\definecolor {salmon3}             {rgb}{0.80,0.44,0.33}
\definecolor {salmon4}             {rgb}{0.55,0.30,0.22}
\definecolor {lightsalmon2}        {rgb}{0.93,0.58,0.45}
\definecolor {lightsalmon3}        {rgb}{0.80,0.51,0.38}
\definecolor {lightsalmon4}        {rgb}{0.55,0.34,0.26}
\definecolor {orange2}             {rgb}{0.93,0.60,0.00}
\definecolor {orange3}             {rgb}{0.80,0.52,0.00}
\definecolor {orange4}             {rgb}{0.55,0.35,0.00}
\definecolor {darkorange1}         {rgb}{1.00,0.50,0.00}
\definecolor {darkorange2}         {rgb}{0.93,0.46,0.00}
\definecolor {darkorange3}         {rgb}{0.80,0.40,0.00}
\definecolor {darkorange4}         {rgb}{0.55,0.27,0.00}
\definecolor {coral1}              {rgb}{1.00,0.45,0.34}
\definecolor {coral2}              {rgb}{0.93,0.42,0.31}
\definecolor {coral3}              {rgb}{0.80,0.36,0.27}
\definecolor {coral4}              {rgb}{0.55,0.24,0.18}
\definecolor {tomato2}             {rgb}{0.93,0.36,0.26}
\definecolor {tomato3}             {rgb}{0.80,0.31,0.22}
\definecolor {tomato4}             {rgb}{0.55,0.21,0.15}
\definecolor {orangered2}          {rgb}{0.93,0.25,0.00}
\definecolor {orangered3}          {rgb}{0.80,0.22,0.00}
\definecolor {orangered4}          {rgb}{0.55,0.15,0.00}
\definecolor {red2}                {rgb}{0.93,0.00,0.00}
\definecolor {red3}                {rgb}{0.80,0.00,0.00}
\definecolor {red4}                {rgb}{0.55,0.00,0.00}
\definecolor {deeppink2}           {rgb}{0.93,0.07,0.54}
\definecolor {deeppink3}           {rgb}{0.80,0.06,0.46}
\definecolor {deeppink4}           {rgb}{0.55,0.04,0.31}
\definecolor {hotpink1}            {rgb}{1.00,0.43,0.71}
\definecolor {hotpink2}            {rgb}{0.93,0.42,0.65}
\definecolor {hotpink3}            {rgb}{0.80,0.38,0.56}
\definecolor {hotpink4}            {rgb}{0.55,0.23,0.38}
\definecolor {pink1}               {rgb}{1.00,0.71,0.77}
\definecolor {pink2}               {rgb}{0.93,0.66,0.72}
\definecolor {pink3}               {rgb}{0.80,0.57,0.62}
\definecolor {pink4}               {rgb}{0.55,0.39,0.42}
\definecolor {lightpink1}          {rgb}{1.00,0.68,0.73}
\definecolor {lightpink2}          {rgb}{0.93,0.64,0.68}
\definecolor {lightpink3}          {rgb}{0.80,0.55,0.58}
\definecolor {lightpink4}          {rgb}{0.55,0.37,0.40}
\definecolor {palevioletred1}      {rgb}{1.00,0.51,0.67}
\definecolor {palevioletred2}      {rgb}{0.93,0.47,0.62}
\definecolor {palevioletred3}      {rgb}{0.80,0.41,0.54}
\definecolor {palevioletred4}      {rgb}{0.55,0.28,0.36}
\definecolor {maroon1}             {rgb}{1.00,0.20,0.70}
\definecolor {maroon2}             {rgb}{0.93,0.19,0.65}
\definecolor {maroon3}             {rgb}{0.80,0.16,0.56}
\definecolor {maroon4}             {rgb}{0.55,0.11,0.38}
\definecolor {violetred1}          {rgb}{1.00,0.24,0.59}
\definecolor {violetred2}          {rgb}{0.93,0.23,0.55}
\definecolor {violetred3}          {rgb}{0.80,0.20,0.47}
\definecolor {violetred4}          {rgb}{0.55,0.13,0.32}
\definecolor {magenta2}            {rgb}{0.93,0.00,0.93}
\definecolor {magenta3}            {rgb}{0.80,0.00,0.80}
\definecolor {magenta4}            {rgb}{0.55,0.00,0.55}
\definecolor {orchid1}             {rgb}{1.00,0.51,0.98}
\definecolor {orchid2}             {rgb}{0.93,0.48,0.91}
\definecolor {orchid3}             {rgb}{0.80,0.41,0.79}
\definecolor {orchid4}             {rgb}{0.55,0.28,0.54}
\definecolor {plum1}               {rgb}{1.00,0.73,1.00}
\definecolor {plum2}               {rgb}{0.93,0.68,0.93}
\definecolor {plum3}               {rgb}{0.80,0.59,0.80}
\definecolor {plum4}               {rgb}{0.55,0.40,0.55}
\definecolor {mediumorchid1}       {rgb}{0.88,0.40,1.00}
\definecolor {mediumorchid2}       {rgb}{0.82,0.37,0.93}
\definecolor {mediumorchid3}       {rgb}{0.71,0.32,0.80}
\definecolor {mediumorchid4}       {rgb}{0.48,0.22,0.55}
\definecolor {darkorchid1}         {rgb}{0.75,0.24,1.00}
\definecolor {darkorchid2}         {rgb}{0.70,0.23,0.93}
\definecolor {darkorchid3}         {rgb}{0.60,0.20,0.80}
\definecolor {darkorchid4}         {rgb}{0.41,0.13,0.55}
\definecolor {purple1}             {rgb}{0.61,0.19,1.00}
\definecolor {purple2}             {rgb}{0.57,0.17,0.93}
\definecolor {purple3}             {rgb}{0.49,0.15,0.80}
\definecolor {purple4}             {rgb}{0.33,0.10,0.55}
\definecolor {mediumpurple1}       {rgb}{0.67,0.51,1.00}
\definecolor {mediumpurple2}       {rgb}{0.62,0.47,0.93}
\definecolor {mediumpurple3}       {rgb}{0.54,0.41,0.80}
\definecolor {mediumpurple4}       {rgb}{0.36,0.28,0.55}
\definecolor {thistle1}            {rgb}{1.00,0.88,1.00}
\definecolor {thistle2}            {rgb}{0.93,0.82,0.93}
\definecolor {thistle3}            {rgb}{0.80,0.71,0.80}
\definecolor {thistle4}            {rgb}{0.55,0.48,0.55}
\definecolor {gray1}               {rgb}{0.01,0.01,0.01}
\definecolor {gray2}               {rgb}{0.02,0.02,0.02}
\definecolor {gray3}               {rgb}{0.03,0.03,0.03}
\definecolor {gray4}               {rgb}{0.04,0.04,0.04}
\definecolor {gray5}               {rgb}{0.05,0.05,0.05}
\definecolor {gray6}               {rgb}{0.06,0.06,0.06}
\definecolor {gray7}               {rgb}{0.07,0.07,0.07}
\definecolor {gray8}               {rgb}{0.08,0.08,0.08}
\definecolor {gray9}               {rgb}{0.09,0.09,0.09}
\definecolor {gray10}              {rgb}{0.10,0.10,0.10}
\definecolor {gray11}              {rgb}{0.11,0.11,0.11}
\definecolor {gray12}              {rgb}{0.12,0.12,0.12}
\definecolor {gray13}              {rgb}{0.13,0.13,0.13}
\definecolor {gray14}              {rgb}{0.14,0.14,0.14}
\definecolor {gray15}              {rgb}{0.15,0.15,0.15}
\definecolor {gray16}              {rgb}{0.16,0.16,0.16}
\definecolor {gray17}              {rgb}{0.17,0.17,0.17}
\definecolor {gray18}              {rgb}{0.18,0.18,0.18}
\definecolor {gray19}              {rgb}{0.19,0.19,0.19}
\definecolor {gray20}              {rgb}{0.20,0.20,0.20}
\definecolor {gray21}              {rgb}{0.21,0.21,0.21}
\definecolor {gray22}              {rgb}{0.22,0.22,0.22}
\definecolor {gray23}              {rgb}{0.23,0.23,0.23}
\definecolor {gray24}              {rgb}{0.24,0.24,0.24}
\definecolor {gray25}              {rgb}{0.25,0.25,0.25}
\definecolor {gray26}              {rgb}{0.26,0.26,0.26}
\definecolor {gray27}              {rgb}{0.27,0.27,0.27}
\definecolor {gray28}              {rgb}{0.28,0.28,0.28}
\definecolor {gray29}              {rgb}{0.29,0.29,0.29}
\definecolor {gray30}              {rgb}{0.30,0.30,0.30}
\definecolor {gray31}              {rgb}{0.31,0.31,0.31}
\definecolor {gray32}              {rgb}{0.32,0.32,0.32}
\definecolor {gray33}              {rgb}{0.33,0.33,0.33}
\definecolor {gray34}              {rgb}{0.34,0.34,0.34}
\definecolor {gray35}              {rgb}{0.35,0.35,0.35}
\definecolor {gray36}              {rgb}{0.36,0.36,0.36}
\definecolor {gray37}              {rgb}{0.37,0.37,0.37}
\definecolor {gray38}              {rgb}{0.38,0.38,0.38}
\definecolor {gray39}              {rgb}{0.39,0.39,0.39}
\definecolor {gray40}              {rgb}{0.40,0.40,0.40}
\definecolor {gray42}              {rgb}{0.42,0.42,0.42}
\definecolor {gray43}              {rgb}{0.43,0.43,0.43}
\definecolor {gray44}              {rgb}{0.44,0.44,0.44}
\definecolor {gray45}              {rgb}{0.45,0.45,0.45}
\definecolor {gray46}              {rgb}{0.46,0.46,0.46}
\definecolor {gray47}              {rgb}{0.47,0.47,0.47}
\definecolor {gray48}              {rgb}{0.48,0.48,0.48}
\definecolor {gray49}              {rgb}{0.49,0.49,0.49}
\definecolor {gray50}              {rgb}{0.50,0.50,0.50}
\definecolor {gray51}              {rgb}{0.51,0.51,0.51}
\definecolor {gray52}              {rgb}{0.52,0.52,0.52}
\definecolor {gray53}              {rgb}{0.53,0.53,0.53}
\definecolor {gray54}              {rgb}{0.54,0.54,0.54}
\definecolor {gray55}              {rgb}{0.55,0.55,0.55}
\definecolor {gray56}              {rgb}{0.56,0.56,0.56}
\definecolor {gray57}              {rgb}{0.57,0.57,0.57}
\definecolor {gray58}              {rgb}{0.58,0.58,0.58}
\definecolor {gray59}              {rgb}{0.59,0.59,0.59}
\definecolor {gray60}              {rgb}{0.60,0.60,0.60}
\definecolor {gray61}              {rgb}{0.61,0.61,0.61}
\definecolor {gray62}              {rgb}{0.62,0.62,0.62}
\definecolor {gray63}              {rgb}{0.63,0.63,0.63}
\definecolor {gray64}              {rgb}{0.64,0.64,0.64}
\definecolor {gray65}              {rgb}{0.65,0.65,0.65}
\definecolor {gray66}              {rgb}{0.66,0.66,0.66}
\definecolor {gray67}              {rgb}{0.67,0.67,0.67}
\definecolor {gray68}              {rgb}{0.68,0.68,0.68}
\definecolor {gray69}              {rgb}{0.69,0.69,0.69}
\definecolor {gray70}              {rgb}{0.70,0.70,0.70}
\definecolor {gray71}              {rgb}{0.71,0.71,0.71}
\definecolor {gray72}              {rgb}{0.72,0.72,0.72}
\definecolor {gray73}              {rgb}{0.73,0.73,0.73}
\definecolor {gray74}              {rgb}{0.74,0.74,0.74}
\definecolor {gray75}              {rgb}{0.75,0.75,0.75}
\definecolor {gray76}              {rgb}{0.76,0.76,0.76}
\definecolor {gray77}              {rgb}{0.77,0.77,0.77}
\definecolor {gray78}              {rgb}{0.78,0.78,0.78}
\definecolor {gray79}              {rgb}{0.79,0.79,0.79}
\definecolor {gray80}              {rgb}{0.80,0.80,0.80}
\definecolor {gray81}              {rgb}{0.81,0.81,0.81}
\definecolor {gray82}              {rgb}{0.82,0.82,0.82}
\definecolor {gray83}              {rgb}{0.83,0.83,0.83}
\definecolor {gray84}              {rgb}{0.84,0.84,0.84}
\definecolor {gray85}              {rgb}{0.85,0.85,0.85}
\definecolor {gray86}              {rgb}{0.86,0.86,0.86}
\definecolor {gray87}              {rgb}{0.87,0.87,0.87}
\definecolor {gray88}              {rgb}{0.88,0.88,0.88}
\definecolor {gray89}              {rgb}{0.89,0.89,0.89}
\definecolor {gray90}              {rgb}{0.90,0.90,0.90}
\definecolor {gray91}              {rgb}{0.91,0.91,0.91}
\definecolor {gray92}              {rgb}{0.92,0.92,0.92}
\definecolor {gray93}              {rgb}{0.93,0.93,0.93}
\definecolor {gray94}              {rgb}{0.94,0.94,0.94}
\definecolor {gray95}              {rgb}{0.95,0.95,0.95}
\definecolor {gray97}              {rgb}{0.97,0.97,0.97}
\definecolor {gray98}              {rgb}{0.98,0.98,0.98}
\definecolor {gray99}              {rgb}{0.99,0.99,0.99}
\definecolor {darkgrey}            {rgb}{0.66,0.66,0.66}

\newcommand{\new}[1]{{\blue #1}\/}

\newcommand{\resp}[1]{[resp. #1]}

\newcommand{\TODO}[1]{{}}

\newcommand{\ignore}[1]{}

\newcommand{\RSTODO}[1]{{\bf \textcolor{darkgreen}{{\fbox{RS TODO:} #1}}}}
\renewcommand{\RSTODO}[1]{}

 \newcommand{\ignoreinshort}[1]{}
 \newcommand{\ignoreinlong}[1]{{#1}}

\def\makenewenumerate#1#2{%
\newcounter{cnt#1}
\newenvironment{#1}%
{\begin{list}{\makebox[0pt][r]{#2}}%
{\setlength{\itemsep}{0pt}%
 \setlength{\parsep}{.2em}%
 \setlength{\leftmargin}{1.5em}%
 \setlength{\labelwidth}{.4em}%
 \usecounter{cnt#1}}}
{\end{list}}}

\makenewenumerate{myenumerate}{\arabic{cntmyenumerate}.}

\makenewenumerate{renumerate}{\rm(\roman{cntrenumerate})}
\makenewenumerate{renumerateprime}{\rm(\roman{cntrenumerateprime}$'$)}
\makenewenumerate{renumeratesecond}{\rm(\roman{cntrenumeratesecond}$''$)}

\makenewenumerate{aenumerate}{({\it\alph{cntaenumerate}})}
\makenewenumerate{aenumerateprime}{({\it\alph{cntaenumerateprime}$'$})}
\makenewenumerate{aenumeratesecond}{({\it\alph{cntaenumeratesecond}$''$})}

\def\newplaintheorem#1#2{%
\newtheorem{#1plain}{#2}[section]%
\newenvironment{#1}{\begin{#1plain}\rm }{\end{#1plain}}}

\newplaintheorem{notation}{Notation}
\newplaintheorem{cexample}{Counter-Example}

\newcommand{\sref}[1]{\S{}\ref{#1}}
\newcommand{\noi}{\noindent}

\newcommand{\set}[1]{\ensuremath{\{{#1}\}}\xspace}
\newcommand{\imp}{\ensuremath{\rightarrow}\xspace}
\newcommand{\limp}{\ensuremath{\leftarrow}\xspace}
\renewcommand{\iff}{\ensuremath{\leftrightarrow}\xspace}
\newcommand{\defas}{\ensuremath{\stackrel{\text{\tiny def}}{=}}\xspace}
\newcommand{\thus}{\ensuremath{\Longrightarrow}\xspace}

\newcommand{\pos}{\phantom{\neg}}

\newcommand{\trueval}{{\ensuremath{\mathsf{true}}}}
\newcommand{\falseval}{{\ensuremath{\mathsf{false}}}}

\newcommand\mysout{\bgroup \markoverwith{{-}}\ULon}
\newcommand\nosout{\bgroup \markoverwith{{ }}\ULon}
\definecolor{mygray}{rgb}{0.90,0.90,0.90}
\definecolor{mywhite}{rgb}{1.00,1.00,1.00}

\newcommand{\vi}{\ensuremath{\varphi}\xspace}

\renewcommand{\new}[1]{{\em #1}}
\renewcommand{\RSTODO}[1]{{\bf \textcolor{darkgreen}{{\fbox{RS TODO:} #1}}}}
\renewcommand{\paragraph}[1]{\smallskip\noindent{\emph{#1}}}

\newcommand{\alla}{\ensuremath{\underline{\mathbf{A}}}\xspace}
\newcommand{\allb}{\ensuremath{\underline{\mathbf{B}}}\xspace}
\renewcommand{\vi}{\ensuremath{\varphi_{[\alla]}}\xspace}
\newcommand{\vione}{\ensuremath{\varphi'_{[\alla]}}\xspace}
\newcommand{\vitwo}{\ensuremath{\varphi''_{[\alla]}}\xspace}
\renewcommand{\vi}{\ensuremath{\varphi}\xspace}
\renewcommand{\vione}{\ensuremath{\varphi_{1}}\xspace}
\renewcommand{\vitwo}{\ensuremath{\varphi_{2}}\xspace}
\newcommand{\ps}{\ensuremath{\psi_{[\alla,\allb]}}\xspace}

\renewcommand{\trueval}{\ensuremath{\mbox{{\sf T}}\xspace}}
\renewcommand{\falseval}{\ensuremath{\mbox{{\sf F}}\xspace}}
\newcommand{\unknown}{\ensuremath{\mbox{{\sf ?}}\xspace}}

\newcommand{\tval}{\trueval}
\newcommand{\fval}{\falseval}
\newcommand{\uval}{\unknown}

\newcommand{\andmu}{\ensuremath{\bigwedge\!\mu}\xspace}
\newcommand{\andeta}{\ensuremath{\bigwedge\!\eta}\xspace}

\newcommand{\entails}{\models}
\newcommand{\validates}{\vdash}
\renewcommand{\validates}{\Vdash}
\renewcommand{\validates}{\mid\!\approx}

\newcommand{\apply}[2]{\ensuremath{#1|_{#2}}\xspace}
\newcommand{\applymuvi}{\apply{\vi}{\mu}\xspace}
\newcommand{\applyetavi}{\apply{\vi}{\eta}\xspace}

\newcommand{\muof}[1]{\ensuremath{\mu(#1)}}
\newcommand{\etaof}[1]{\ensuremath{\eta(#1)}}

\newcommand{\oksym}{{\Large \ensuremath{\textcolor{darkgreen}{\checkmark}}}\xspace}
\newcommand{\noksym}{{\LARGE \ensuremath{\textcolor{red}{\mathbf{\times}}}}\xspace}

\newcommand{\shannon}[2]{\ensuremath{#1|_{\exists#2}}\xspace}
\renewcommand{\shannon}[2]{\ensuremath{{\sf SE}[{\exists#2}.#1}]\xspace}
\newcommand{\shannonpsib}{\shannon{\psi}{\allb}}

\newcommand{\predabsphi}{\ensuremath{{\sf PredAbs}(\phi,\set{\phi_i}_{i=1}^N)}}
\renewcommand{\predabsphi}{\ensuremath{{\sf PredAbs}(\phi,{\mathbf \Phi})}}

\newcommand{\verbvalidate}{validate}
\newcommand{\verbvalidated}{validated}
\newcommand{\verbvalidates}{validates}
\newcommand{\verbvalidating}{validating}
\newcommand{\verbvalidation}{validation}

\renewcommand{\verbvalidate}{evaluate to true}
\renewcommand{\verbvalidated}{evaluated to true}
\renewcommand{\verbvalidates}{evaluates to true}
\renewcommand{\verbvalidating}{evaluating to true}
\renewcommand{\verbvalidation}{evaluation to true}

\renewcommand{\RSTODO}[1]{\todo[size=\tiny,color=green!40]{{\small{#1}}}}

\newcommand{\longversion}{true}
\ifthenelse{\equal{\longversion}{true}}
{%
  \renewcommand{\ignoreinshort}[1]{#1}
  \renewcommand{\ignoreinlong}[1]{}
  \specialcomment{IGNOREINSHORT}{\begingroup}{\endgroup}
  \excludecomment{IGNOREINLONG}
}%
{%
 \renewcommand{\ignoreinshort}[1]{}
 \renewcommand{\ignoreinlong}[1]{\textcolor{midnightblue}{#1}}
 \excludecomment{IGNOREINSHORT}
  \specialcomment{IGNOREINLONG}{\begingroup\color{black}}{\endgroup}
}
 \excludecomment{IGNORE}

\excludecomment{TOBEDROPPED}

\renewcommand{\TODO}[1]{\todo[inline,color=green!40]{{\small{#1}}}}

\begin{document}

\pagestyle{plain}
\pagenumbering{roman}

\pagestyle{plain}
\pagenumbering{arabic}

\title{%
Are You Satisfied by This Partial Assignment?
%
%
}

\author{
Roberto Sebastiani 
}

\institute{%
DISI, University of Trento, Italy%
}

\maketitle
 \begin{abstract}
Many procedures for SAT and 
SAT-related problems --in particular for those requiring the complete
enumeration of satisfying truth assignments--
rely their efficiency
on the detection of {\em partial} assignments satisfying an input
formula.
In this paper we analyze the notion of partial-assignment
satisfiability --in particular when dealing with non-CNF and
existentially-quantified formulas-- raising a flag about the
ambiguities and subtleties of this concept, and investigating their
practical consequences.
This may drive the development of more effective assignment-enumeration algorithms.

%

 \end{abstract}

\section{Motivations and Goals}
\label{sec:intro}
Many procedures for SAT and SAT-related problems (e.g.  Analytic
Tableaux \cite{smullyan1}, DPLL \cite{davis7}, OBDDs \cite{bryant2})
rely their efficiency on the detection of {\em partial} truth
assignments $\mu$ satisfying an input propositional formula $\vi$,
which allows to state that (i) $\vi$ is satisfiable and (ii) all total
assignments extending $\mu$ satisfy \vi.
In particular,
when it comes to SAT-based problems requiring
the {\em complete enumeration} of satisfying assignments (e.g. \#SAT
\cite{GSS09HBSAT}, Lazy SMT \cite{BSST09HBSAT}, AllSAT and AllSMT
\cite{allsmt}, satisfiability of modal and description logics
\cite{ST09HBSAT}, Weighted Model Integration \cite{MorPasSeb17}),
the ability of enumerating satisfying {partial} assignments which are
as small as possible is essential, because
each of them avoids the enumeration of the whole subtree of total
assignments extending it, whose size is exponential in the number of
unassigned propositions.

In this paper we 
analyze the notion of partial-assignment
satisfiability  --in particular when dealing with  non-CNF and
existentially-quantified formulas-- 
raising  a flag about the ambiguities and subtleties
of this concept, and
investigating their practical consequences.
We notice, analyze and discuss the following facts.\footnote{Not necessarily in this order.}

First, despite its widespread (implicit) usage 
in algorithms,
there seems to be no general and universally-agreed notion of
partial-assignment satisfaction.
Most authors do not
define partial-assignment satisfaction explicitly, or define it only when dealing with
(tautology-free) CNF formulas (e.g.  \cite{KBK09HBSAT}).
We stress the fact that this is not simply an issue of
the meaning of the word ``satisfy'': regardless which ``\verb|verb|'' one
might  use for it (e.g. ``satisfy'', ``entail'', ``imply'', ``evaluate to
true'',...), we would like a universally-agreed criterion
to establishing that, if a partial truth-assignment ``\verb|verb|''s a formula
$\vi$, then (i) $\vi$ is satisfiable and (ii) all total truth
 assignments extending it satisfy $\vi$. 

Second, for (tautology-free) CNF formulas the sentence
``{\em a partial truth assignment $\mu$ satisfies a formula $\vi$}''
may be indifferently be interpreted either as 
``$\mu$ {\em \verbvalidates{}} $\vi$''
(i.e. ``applying $\mu$ to $\vi$ makes $\vi$ true'')
or as
``$\mu$ {\em entails} $\vi$''
(i.e. ``all total assignments extending $\mu$
satisfy $\vi$'')
because in this case the two  concepts are equivalent.
Consequently, satisfiability and enumeration algorithms for CNF
formulas typically use \verbvalidation{} as criterion to conclude that
the current partial assignment satisfies the input formula, because it
is much cheaper and easier to implement than entailment.

Third, and most importantly,
 %
{\em for non-CNF formulas
\verbvalidation{} is strictly stronger than entailment}, and they have
complementary properties.
Consequently, whereas using \verbvalidation{} as partial-assignment
satisfiability
criterion is much cheaper and easier to implement,
adopting entailment allows for detecting satisfiability earlier and
thus for producing smaller partial truth assignments.
We also show that, 
whereas equivalent 
  formulas are always entailed by the
same partial assignments, this is not the case for
\verbvalidation{}, that is, {\em equivalent formulas are not always
  \verbvalidated{} by the same partial assignments}.
This would be an embarrassing fact
if we adopted \verbvalidation{} as the definition of  partial-assignment
satisfiability for non-CNF formulas.
%
We remark that standard Tseitin-style CNF-ization does not solve these
issues, because
it may loose 
information regarding partial-assignment  \verbvalidation{} or entailment.

Fourth,
the same issues apply also  for
existentially-quantified formulas, even CNF ones.
%
This is very important, because in many application domains,
fundamental operations ---like {\em  
  preimage computation} in symbolic model checking
(see e.g. \cite{burch1}) or {\em predicate abstraction} in SW verification
(see e.g. \cite{graf_predabs97,beyercgks09})---
require dealing with existentially-quantified formulas and with
the enumeration of partial assignments ``satisfying'' them.


Fifth, different algorithms handling non-CNF formulas implicitly implement different notions of
partial-assignment satisfaction.
%
E.g.,
Analytic Tableaux \cite{smullyan1} and
(non-CNF) DPLL \cite{davis7} implicitly enumerate partial assignment
{\verbvalidating{}} the input formulas,
whereas 
OBDDs \cite{bryant2} implicitly enumerate partial assignment {entailing}
  them. 
Also, e.g., techniques like {\em pure-literal
filtering} 
\cite{sebastiani07,BSST09HBSAT} in lazy
SMT 
implicitly aim at reducing a
total assignment to a partial one {\verbvalidating{}} the input formula.

\begin{TOBEDROPPED}
Surprisingly,
there seems to be no general and universally-agreed definition of
partial-assignment satisfaction:
most authors do not
define partial-assignment satisfaction explicitly, or define it only when dealing with
(tautology-free) CNF formulas (e.g.  \cite{KBK09HBSAT});
%
also, different algorithms implicitly implement different notions of
partial-assignment satisfaction (see below).
%

We stress the fact that this is not simply an issue of
the meaning of the word ``satisfy'': regardless which ``verb'' one
might  use for it (e.g. ``satisfy'', ``entail'', ``imply'', ``evaluate to
true'',...), we would like a universally-agreed criterion
to establishing that, if a partial truth-assignment ``verbs'' a formula
$\vi$, then (i) $\vi$ is satisfiable and (ii) all total truth
 assignments extending it satisfy $\vi$. 
\ignore{
we need a universally-agreed criterion
 complying with the two conditions (i) and (ii)
above.
} 




\RSTODO{move up?}
In this paper we want to address this issue explicitly. We
analyze the notion of partial-assignment
satisfiability,
raising  a flag about the ambiguities and subtleties
of this concept  --in particular when dealing with  non-CNF and
existentially-quantified formulas-- 
and we investigate their practical consequences.

We start by noticing the following fact:
whereas for (tautology-free) CNF formulas
``{\em a partial truth assignment $\mu$ satisfies a formula $\vi$}''
may be indifferently be interpreted either as 
``$\mu$ {\em \verbvalidates{}} $\vi$''
(i.e. ``applying $\mu$ to $\vi$ makes $\vi$ true'')
or as
``$\mu$ {\em entails} $\vi$''
(i.e. ``all total assignments extending $\mu$
satisfy $\vi$'')
because in this case the two  concepts are equivalent,
 for non-CNF formulas instead --and for
existentially-quantified formulas, even CNF ones--
{\em \verbvalidation{} is strictly stronger than entailment}, and they have
complementary properties. 
\ignore{
First we raise a flag on the fact that by
``{\em a partial truth assignment $\mu$ satisfies a formula $\vi$}'',
we may mean either that $\mu$ {\em entails} $\vi$
(i.e. that all total assignments extending $\mu$
satisfy $\vi$) or that $\mu$ {\em \verbvalidates{}} $\vi$
 (i.e. that applying $\mu$ to $\vi$ makes $\vi$ true).
Typically most authors in the SAT community
leave this definition
implicit, or provide a definition only for (tautology-free) CNF
formulas (e.g. \cite{KBK09HBSAT}).
%
We notice that, whereas for (tautology-free) CNF formulas entailment 
and \verbvalidation{} are equivalent, for non-CNF formulas --and for
existentially-quantified formulas, even CNF ones--
{\em \verbvalidation{} is strictly stronger than entailment}, and they have
complementary properties. 
}
We also notice that standard
 (Tseitin-style) CNF-ization does not solve this
issue, because 
it may lose 
information regarding partial-assignment entailment or \verbvalidation{}.
\end{TOBEDROPPED}


\smallskip
Overall, the theoretical considerations above suggest to adopt entailment
  as general definition of partial-assignment satisfiability, although
  \verbvalidation{} is a cheaper though less-effective criterion which can 
  (most) often be adopted in
  actual implementations. 
%
%
  However, since partial assignments entailing $\vi$ are in general
  subsets of those \verbvalidating{} $\vi$, using entailment rather
  than \verbvalidation{} as satisfiability criterion allows for
  producing smaller partial assignments, and hence possibly
  drastically reducing their number, in particular in the presence of
  existentially-quantified formulas.  This may drive the development
  of more effective assignment-enumeration algorithms.
%

\ignore{
\paragraph{A poll.}
In order to support the analysis, I've set up an email
poll among some current or past SAT PC and some SMT PC members, asking
for the truth/falsehood of five statements on the
notion of partial-assignment satisfiability.
The result are available at \cite{poll} %
 and as an appendix submitted with this paper. 
Although only 20 out of 65 people replied, 
the result 
reveals the lack of general agreement among the repliers on the notion of
partial-assignment satisfiability and on its consequences: e.g.,
the repliers split evenly into supporters 
of entailment, of \verbvalidation{}, and people considering them equivalent.
}

\paragraph{Motivation.} The analysis presented in this paper was
triggered
by the effort of conceiving more efficient procedures for predicate
abstraction in SMT for improving Weighted Model Integration
\cite{MorPasSeb17,morettin_aij19},
which forced me to elaborate on the distinction between 
\verbvalidation{} and entailment. Before then, I personally used to see
partial-assignment satisfiability as {entailment}
 (see \cite{gs-infocomp2000}) without paying attention to this distinction.

{\paragraph{Content.} The rest of the paper is organized as follows.
\sref{sec:background} provides the necessary notation, terminology
and concepts used in the paper.
\sref{sec:partialsat} introduces \verbvalidation{} and entailment
 for generic propositional formulas and discusses their
relative properties and use.
\sref{sec:partialsat-exist} lifts the discussion to
existentially-quantified formulas.
\sref{sec:concl} provides some conclusions and future-work suggestions.
}

\section{Background}
\label{sec:background}

In this section we introduce the 
notation and terminology adopted in this paper.
Moreover, in order to avoid any ambiguity
(although at the risk of being a little pedantic),
we recall the standard
syntax and semantics of propositional logics, plus some basic facts.   

%

\paragraph{Notation.}
In what follows
$\tval$, $\fval$, $\uval$ denote the
truth values ``true'', ``false'' and ``unknown'' respectively;
$\top$, $\bot$ denote the logic constants
``true'' and ``false'' respectively; 
%
$A$, $B$ denote propositional atoms;
%
$\varphi,\phi,\psi$ denote propositional formulas;
%
$\mu,\eta,\gamma$ denote truth value assignments.
%
The symbols $\alla\defas\set{A_1,...,A_N}$ and
$\allb\defas\set{B_1,...,B_K}$ 
denote disjoint sets of propositional atoms. 
More precisely, $\vi$, $\phi$ and $\psi$ denote generic propositional formulas
built on $\alla$, $\allb$ and $\alla\cup\allb$ respectively;
 $\eta$ and $\mu$ denote total and a partial
assignments on $\alla$ respectively;
$\delta$ denote total assignments on $\allb$. (All above symbols may possibly
have subscripts).

\paragraph{Syntax.}
A \new{propositional formula} is defined inductively as follows: 
the constants $\top$ and $\bot$ (denoting the truth values true and false)
are formulas; 
   a {propositional atom} 
   $A_1,A_2,A_3,...$ is a formula;
   if $\varphi_1$ and $\varphi_2$ are formulas, then 
\ignore{ $\neg\varphi_1$ and 
  $\varphi_1\bowtie\varphi_2$ are formulas, s.t. $\bowtie\ \in\set{\wedge,\vee,\imp,\limp,\iff}$.
}
  $\neg\varphi_1$ and 
  $\varphi_1\wedge\varphi_2$
%
  are formulas.
We use the standard Boolean abbreviations:  
``$\varphi_1\vee\varphi_2$'' for  
``$\neg(\neg\varphi_1\wedge\neg\varphi_2)$'', 
``$\varphi_1\imp\varphi_2$'' for ``$\neg(\varphi_1\wedge\neg\varphi_2)$'', 
``$\varphi_1\iff\varphi_2$'' for 
``$\neg(\varphi_1\wedge\neg\varphi_2) \wedge
\neg(\varphi_2\wedge\neg\varphi_1)$''.
%
A \new{literal} is either an atom (a \new{positive
 literal}) or its negation (a \new{negative
 literal}).
(If $l$ is a negative literal $\neg A_i$,
then by ``$\neg l$'' we conventionally mean $A_i$ rather than
$\neg\neg A_i$.)
A \new{clause} is a disjunction of literals $\bigvee_j l_j$.
A \new{cube} is a conjunction of literals $\bigwedge_j l_j$.
$\varphi$ is in \new{Conjunctive Normal Form (CNF)} iff it is
   a conjunction of clauses:
$
\bigwedge_{i=1}^{L}\bigvee_{j_i=1}^{K_i} l_{j_i}
$.

\paragraph{Semantics.}
Given $\alla\defas\set{A_1,...,A_N}$,
a map  $\eta:\alla\longmapsto\set{\trueval,\falseval}^N$ is a
\new{total truth assignment} for $\alla$. 
\ignoreinshort{We assume $\eta(\top)\defas \trueval$ and $\eta(\bot)\defas \falseval$.}
We  represent $\eta$ as a \new{set of literals}
$\eta\defas \set{A_i\ |\ \eta(A_i)=\trueval}\cup\set{\neg A_i\ |\
  \eta(A_i)=\falseval}$. We sometimes represent $\eta$ also as a
\new{cube} 
$
\bigwedge_{\eta(A_i)=\trueval} A_i \wedge 
\bigwedge_{\eta(A_i)=\falseval}\neg A_i
$ which we denote as ``$\andeta$'' so that to distinguish the set and the cube
representations. 

%
\begin{figure}[t!]
  \small
  \begin{tabular}{cc}
  \begin{minipage}[t]{0.49\textwidth}
    $
      \begin{array}{||c||l|l|l|l|l|l|l|l|l||}
  \hline    
  \muof{\vi_1}            & \tval & \tval & \tval & \uval & \uval & \uval &\fval & \fval & \fval\\
  \muof{\vi_2}            & \tval & \uval & \fval & \tval & \uval & \fval &\tval & \uval & \fval\\
     \hline
  \muof{\neg\vi_1}        & \fval & \fval & \fval & \uval & \uval & \uval &\tval & \tval & \tval\\
  \muof{\vi_1\wedge\vi_2} & \tval & \uval & \fval & \uval & \uval & \fval &\fval & \fval & \fval\\
  \muof{\vi_1\vee\vi_2}   & \tval & \tval & \tval & \tval & \uval & \uval &\tval & \uval & \fval\\
  \muof{\vi_1\imp\vi_2}   & \tval & \uval & \fval & \tval & \uval & \uval &\tval & \tval & \tval\\
  \muof{\vi_1\iff\vi_2}   & \tval & \uval & \fval & \uval & \uval & \uval &\fval & \uval & \tval\\
     \hline
      \end{array}
      $
\caption{\label{fig:threeval}
Three-value-semantics of $\muof{\vi}$ in terms of \set{\tval,\fval,\uval}
(``true'', ``false'', ``unknown''). \newline
\ignoreinlong{The definition of
$\muof{\vi_1\bowtie\vi_2}$ s.t. $\bowtie\ \in\set{\vee,\imp,\limp,\iff}$
follows straighforwardly.}
}
  \end{minipage}
      &
  \begin{minipage}[t]{0.4\textwidth}
    $
    \begin{array}{|cl|cl|}
  \hline    
      \neg\top & \Rightarrow \bot & \neg \bot & \Rightarrow \top \\
      \top\wedge\vi, \vi\wedge\top  & \Rightarrow \vi & \bot\wedge\vi,  \vi\wedge\bot& \Rightarrow \bot \\
      \top\vee\vi, \vi\vee\top& \Rightarrow \top & \bot\vee\vi, \vi\vee\bot& \Rightarrow \vi \\
      \top\imp\vi & \Rightarrow \vi & \bot\imp\vi & \Rightarrow \top \\
      \vi\imp\top & \Rightarrow \top & \vi\imp\bot & \Rightarrow \neg\vi \\
      \top\iff\vi, \vi\iff\top& \Rightarrow \vi & \bot\iff\vi, \vi\iff\bot& \Rightarrow \neg\vi \\
      & & &  \\
  \hline         
    \end{array}
    $\\ 
    \caption{\label{fig:boolprop}
    Propagation of truth values 
through the Boolean connectives.}
\end{minipage}%
\end{tabular}
\end{figure}

A map  $\mu:\alla'\longmapsto\set{\trueval,\falseval}^{N'}$
s.t. $\alla'\subseteq\alla$ and $N'\defas||\alla'||$ is a
\new{partial truth assignment} for $\alla$. 
%
%
As with total assignments, we can represent $\mu$ as a set of literals or as
a cube, denoted with ``$\andmu$''. 
Using a three-value logic we extend $\mu$ to \alla{} as
$\mu:\alla\longmapsto\set{\trueval,\falseval,\unknown}^{N}$
by assigning to \unknown{} (unknown) the unassigned atoms in $\alla\setminus \alla'$.
Then we extend the semantics of $\mu$ to any formula $\vi$ on \alla as
described in Figure~\ref{fig:threeval}. We say that $\mu$
{\em evaluates to true \resp{false}} $\vi$ if $\muof{\vi}=\tval$ \resp{$\muof{\vi}=\fval$}. 

By ``{\em apply a partial assignment $\mu$ to $\varphi$}'' we mean 
``substitute all instances of each assigned
$A_i$ in $\varphi$ with the truth value in $\set{\top,\bot}$  assigned by
$\mu$ and then apply recursively the standard
propagation of truth values through the Boolean connectives
described in Figure~\ref{fig:boolprop}.
%
We denote by ``$\applymuvi$'' (``{\em residual of $\vi$ under $\mu$''}) the formula resulting from applying
$\mu$ to $\vi$.
The following fact follows straightforwardly. 

\begin{property}
  \label{prop:eval-simplifies}
$\applymuvi$ is $\top$ iff $\muof{\vi}=\tval$ and
$\applymuvi$ is $\bot$ iff $\muof{\vi}=\fval$. 
\end{property}

\noindent
Notice that total assignments are a subcase of partial ones, so that all
above definitions and facts apply also to total assignments $\eta$.

Given a total truth assignment $\eta$ on \alla{} and some formulas
$\vi,\vi_1,\vi_2$ on \alla, the sentence ``$\eta$ {\em satisfies} $\varphi$'', written
``$\eta\models\varphi$'', 
is defined recursively on the structure of $\varphi$ as follows:
$\eta\models \top$, 
$\eta\not\models \bot$, 
$\eta\models A_i$ if and only if $\eta(A_i)=\trueval$,
$\eta\models\neg\vione$ if and only if $\eta\not\models\vione$,
$\eta\models\vione\wedge\vitwo$ if and only if
$\eta\models\vione$ and $\eta\models\vitwo$.
{(The definition of $\eta\models\vi_1\bowtie\vi_2$ for the other
  connectives  follows straightforwardly from their definition in terms
  of $\neg,\wedge$.)}
%
$\varphi$ is \new{satisfiable} iff $\eta\models\varphi$ for some total truth
assignment $\eta$ on $\alla$.
$\varphi$ is \new{valid} (written ``$\models \vi$'') iff $\eta\models\varphi$ for every total truth
assignment $\eta$ on $\alla$. 
\new{$\varphi_1$ entails $\varphi_2$} (written \new{``$\varphi_1\entails
     \varphi_2$''}) iff,  for every total assignment $\eta$ on
   $\alla$, if
 $\eta\models\varphi_1$
then $\eta\models\varphi_2$.
$\varphi_1$ and $\varphi_2$ are \new{ equivalent} iff 
$\varphi_1\entails  \varphi_2$ and $\varphi_2\entails  \varphi_1$.
Consequently: $\varphi$ is unsatisfiable iff $\neg\varphi$ is valid;
$\vione\models\vitwo$ iff $\vione\imp\vitwo$ is valid;
a clause $\bigvee_i l_i$ is valid (aka is a \new{tautology}) iff both $A_i$ and $\neg A_i$
occur in it for some $A_i$; a CNF formula \vi is valid iff either it
is $\top$ or all its clauses are tautologies. 
We say that a CNF formula is \new{tautology-free} iff none of its clauses
is a tautology.
\ignore{
Hereafter $\vi$ denotes a propositional formula
built on $\alla$, $\eta$ and $\mu$ denote a total and a partial
assignment on $\alla$ respectively (all possibly with subscripts). 
}

\ignore{The following facts on total assignments follow directly from the
  above definitions.}
The following facts follow straightforwardly and are of interest for our discussion.

\begin{property}
\label{prop:total-properties}
Let $\eta$ be a total truth assignment on \alla{} and
$\vi,\vione,\vitwo$ be formulas on \alla. 
\begin{renumerate}
\item 
$\eta\models\varphi$ iff $\andeta\entails\varphi$.
\item 
If $\vione$ and $\vitwo$ are  equivalent, then 
$\eta\models\vione$ iff $\eta\models\vitwo$.
\item
$\eta\models\varphi$ iff \applyetavi{} is $\top$ (also, by
Property~\ref{prop:eval-simplifies}, iff $\etaof{\vi}= \tval$). 
\item 
Checking
if $\eta$ satisfies $\vi$
requires at most a polynomial amounts of steps.  
\end{renumerate}  
\end{property}
\noindent
\ignoreinshort{Notice that Property~\ref{prop:total-properties}(i) justifies the
  usage of ``$\models$'' 
for both satisfiability and entailment.}

\paragraph{CNF-ization.}
Every generic formula $\vi$ on \alla can be encoded into a CNF formula $\psi$
on $\alla\cup\allb$ for some \allb by applying (variants of) Tseitin
CNF-ization \cite{tseitin1}, consisting e.g. in applying recursively bottom-up 
the rewriting rule: 
\begin{eqnarray}
\vi &\Rightarrow&
\vi[(l_{j1} \bowtie l_{j2})\mapsto B_j]\wedge CNF(B_j \iff (l_{j1} \bowtie l_{j2}))  
\end{eqnarray}
until the resulting formula $\psi$ is in CNF, where $l_{j1},l_{j2}$ are
literals, $\bowtie\ \in\ \set{\wedge,\vee,\imp,\limp,\iff}$ and 
$CNF()$ is the validity-preserving CNF conversion based on DeMorgan
rules (e.g., 
$CNF(B \iff (l_1 \wedge l_2)) \defas 
(\neg B \vee l_1) \wedge 
(\neg B \vee l_2) \wedge 
(B \vee \neg l_1 \vee \neg l_2)$).
$\psi$ is s.t. $\eta\models\vi$ iff exists a total assignment $\delta$ on
\allb s.t. $\eta\cup\delta\models\psi$, and the size of $\psi$ is
linear wrt. that of $\vi$.

\paragraph{Existentially-quantified formulas.}
%
A total truth assignment $\eta$ satisfies $\exists \allb. \psi$,
written ``$\eta\models\exists \allb. \psi$'',
iff exists a total truth assignment $\delta$ on \allb s.t.
$\eta\cup\delta\models\psi$.
We call the {\em Shannon expansion} \shannonpsib of the existentially-quantified
formula $\exists \allb. \psi$ the propositional formula on \alla
defined as 
\begin{eqnarray}
\label{eq:shannon}
\shannonpsib\defas 
\bigvee_{\delta_i\in \set{\top,\bot}^K}
\apply{\psi}{\delta_i}
\end{eqnarray}
{Notice that some $\apply{\psi}{\delta_i}$ may be inconsistent or $\bot$.}
The following property derives directly from the above definitions.

\begin{property}
\label{prop:total-shannon}
Let   $\psi$ be a formula on
$\alla\cup\allb$ and $\eta$ be a total truth assignment on \alla.
Then\\
  $\eta\models\exists \allb. \psi$\ iff\ $\eta\models\shannonpsib$. 
\end{property}

\section{Partial-assignment satisfiability for propositional formulas}
\label{sec:partialsat}
\ignore{What do we mean by ``a \emph{partial} assignment $\mu$ satisfies $\vi$"?}
We wish to provide a satisfactory definition of partial-assignment
satisfiability for a generic propositional formula ---i.e., non
necessarily (tautology-free) CNF. \RSTODO{Cambiare narrazione. Unfortunately...}

One first possibility is to see partial-assignment satisfiability as
{\em \verbvalidation}. 
\begin{definition}
\label{def:validation}
We say that a {\em partial} truth assignment $\mu$  {\bf \verbvalidates{}} $\varphi$ 
iff $\muof{\vi}=\tval$ (or, equivalently by
Property~\ref{prop:eval-simplifies}, iff $\apply{\mu}{\vi}=\top$).
We denote this fact with ``$\mu\validates\varphi$''.
\end{definition}
One second possibility is to see partial-assignment satisfiability as
{\em entailment}. 

\begin{definition}
\label{def:entailment}
  We say that a {\em partial} truth assignment $\mu$  {\bf entails} $\varphi$
if and only if, for every total truth assignments $\eta$
s.t.$\mu\subseteq\eta$, 
$\eta$ satisfies $\varphi$. 
We denote this fact with ``$\mu\entails\varphi$''.
\end{definition}

\noindent
Notice that both \verbvalidation{} and entailment are {\em semantic}
definitions.
Due to Property~\ref{prop:eval-simplifies}, \verbvalidation{}
has also an easy-to-check
syntactic characterization as ``$\apply{\mu}{\vi}=\top$''. 

In substance,
Definition~\ref{def:validation} extends to partial
assignments 
Property~\ref{prop:total-properties}(iii),
whereas 
Definition~\ref{def:entailment} extends to partial
assignments Property~\ref{prop:total-properties}(i).
Ideally, a suitable definition of partial-assignment satisfiability
should verify all statements in Property~\ref{prop:total-properties},
in particular (ii) and (iv). 
In practice, unfortunately, at least for generic (non-CNF) formulas, we see
 this is not the case. 

When the formula $\vi$ is in CNF and does not contain
valid clauses --which however are easy to eliminate by preprocessing-- 
then Definitions~\ref{def:validation} and
\ref{def:entailment}  are equivalent: 
$\mu\validates\varphi$ iff $\mu\entails\varphi$.
In fact, if $\mu\validates\vi$ then,  for every $\eta$ s.t.
$\eta\supseteq\mu$, $\eta\validates\vi$ and thus $\eta\models\vi$, hence
$\mu\entails\vi$;
also, if $\mu\entails\vi$ then $\applymuvi$ is a valid
  CNF formula which does not contain valid clauses, so that $\applymuvi$
  must be $\top$, hence $\mu\validates\vi$.

Unfortunately, when dealing with generic (non-CNF) formulas, 
we notice that Definitions~\ref{def:validation} and
\ref{def:entailment}  are \emph{not equivalent}, the former being
strictly stronger than the latter. 
In fact, as above, if $\mu\validates\vi$ then
$\mu\entails\vi$, whereas the converse is not true: e.g.,
if $\vi\defas(A_1\wedge A_2)\vee (A_1\wedge \neg A_2)$ and 
$\mu\defas\set{A_1}$, 
then 
$\mu\entails\vi$ but $\mu\not\validates\vi$.
This leads to the following statement.

\begin{proposition}
\label{prop:stronger}
If a partial truth assignment $\mu$ {\verbvalidates{}}
$\varphi$, then it also {entails} $\varphi$, but the converse does
not hold.
\end{proposition}
\begin{IGNOREINSHORT}
\begin{example}
Let
$\mu\defas\set{A_1,..,A_M}$ s.t. $M<N$ and
$\vi\defas\bigvee_i (A_i\wedge cube_i)$ s.t. each $cube_i$ is a cube and $\bigvee_i cube_i$ is
valid and does not contain occurrences of the atoms $A_1,..,A_M$. Then 
$\mu\entails\vi$ but \applymuvi{} is the valid formula $\bigvee_i
cube_i$, so that $\mu\not\validates\vi$. \hfill $\diamond$
\end{example}  
\end{IGNOREINSHORT}

We try to build a counterpart of Property~\ref{prop:total-properties}
for Definitions~\ref{def:validation} and
\ref{def:entailment} respectively, but in both cases we fail to
achieve all points (i)-(iv) in Property~\ref{prop:total-properties},
resulting into complementary situations.
%

From Definition~\ref{def:validation}
we easily derive the following. 
(Here ``\oksym{}'' \resp{``\noksym{}''} denotes facts from
Property~\ref{prop:total-properties} which are \resp{are not} preserved.)

\begin{property}
\label{prop:partial-validation}
Let $\mu$ be a partial truth assignment on \alla{} and
$\vi,\vione,\vitwo$ be formulas on \alla. 
\begin{renumerate}
\item \noksym{}
If $\mu\validates\varphi$ then $\andmu\entails\varphi$, {but not vice versa}.
\item \noksym{}
If $\vione$ and $\vitwo$ are  equivalent, this {does not} imply
that $\mu\validates\vione$ iff $\mu\validates\vitwo$.
\item \oksym{}
  $\mu\validates\varphi$ iff \applymuvi{} is $\top$
     (also, iff $\muof{\vi}= \tval$ by Property~\ref{prop:eval-simplifies}). 

\item \oksym{}
Checking
if $\mu\validates\vi$  requires at most a polynomial amount of steps.
\end{renumerate}  
\end{property}

From Definition~\ref{def:entailment}
we easily derive the following. 

\begin{property}
\label{prop:partial-entailment}
Let $\mu$ be a partial truth assignment on \alla{} and
$\vi,\vione,\vitwo$ be formulas on \alla. 
\begin{renumerate}
\item \oksym{}
$\mu\models\varphi$ iff $\andmu\entails\varphi$.
\item \oksym{}
If $\vione$ and $\vitwo$ are  equivalent, then 
$\mu\models\vione$ iff $\mu\models\vitwo$.
\item \noksym{}
  $\mu\models\varphi$ iff \applymuvi{} is a valid formula, not
  necessarily $\top$
   (also, in general $\muof{\vi} \neq \tval$). 
\item \noksym{}
Checking
if $\mu\entails\vi$ is \emph{co-NP-complete}.
\footnote{In fact, checking the validity of \vi{}
translates into verifying that the empty assignment
entails it.}
\end{renumerate}  
\end{property}
\noindent

On the one hand, 
the advantage of adopting \verbvalidation{} for checking
partial-assignment satisfiability is that it matches the intuition
and practical need that the process of checking it should be fast (Property~\ref{prop:partial-validation}(iv)).  
On the other hand, 
the main drawback is that 
that equivalent although syntactically different 
formulas may be satisfied by different sets of partial
assignments  (Property~\ref{prop:partial-validation}(ii)), 
which looks theoretically awkward. 

\TODO{We believe this is quite embarassing...}

On the one hand, 
the advantage of adopting entailment
for checking
partial-assignment satisfiability is that it matches the intuition 
and theoretical requirement
that equivalent formulas should be satisfied by the same assignments,
even partial ones (Property~\ref{prop:partial-entailment}(ii)). 
On the other hand, the price to pay is that the resulting problem 
is co-NP-complete  (Property~\ref{prop:partial-entailment}(iv)),
because it is equivalent to checking the validity of the residual \applymuvi.

\smallskip
Due to Proposition~\ref{prop:stronger}, every
 partial assignments entailing the input formula is  a \emph{subset}
 of some other(s) \verbvalidating{} it.
Therefore, for an assignment-enumeration algorithm,
 being able to enumerate partial 
assignments entailing the input formula rather than simply \verbvalidating{}
it  may (even drastically) reduce the number of the satisfying assignment
enumerated. 

For instance we notice that, when applied to satisfiable formulas,
OBDDs \cite{bryant2} produce branches representing
partial assignments which {\em entail} the input formula
(Definition~\ref{def:entailment}), because if $\mu\entails\vi$ then
\applymuvi is
valid (Property~\ref{prop:partial-entailment}(iii)), so that its corresponding 
sub-OBDD is reduced into the $\top$ node.
Instead SAT/AllSAT algorithms like \new{Analytic Tableaux} 
\footnote{Notice that Analytic Tableaux may generate duplicated
 or subsumed  assignments (see \cite{dagostino1,gs-infocomp2000})}
\cite{smullyan1} 
or ``classic'' \new{DPLL}~%
\footnote{Classic DPLL procedure \cite{davis7} was
  designed to work for CNF  
  formulas. Nevertheless it is easy to produce non-CNF a version of this
procedure (see e.g. \cite{armando5}).} 
\cite{davis7} produce branches representing
partial assignments which {\em \verbvalidate} the input formula
(Definition~\ref{def:validation}) because, unlike with OBDDs,
as soon as it is produced (a branch corresponding to) an assignment
$\mu$ s.t. 
$\mu\entails\vi$ but $\mu\not\validates\vi$, they do not realize it and
proceed the search until they extend it to some $\mu'\supset\mu$ s.t. 
$\mu'\validates\vi$, extending the search tree of up to
$2^{||\mu'||-||\mu||}$ branches.  
%

\newcommand{\mainwffone}{\ensuremath{(A_1\wedge  A_2)\vee(A_1\wedge \neg A_2)}}
\newcommand{\mainwfftwo}{\ensuremath{(\neg A_3\wedge  A_4)\vee(\neg A_3\wedge \neg A_4)}}
\newcommand{\mainwff}{\ensuremath{(\mainwffone)\wedge(\mainwfftwo)}}
\newcommand{\mainwffstacked}{\ensuremath{
    \begin{array}{c}
      \mainwffone \\
      \mainwfftwo
    \end{array}}}
\newcommand{\wffl}{\ensuremath{(A_1\wedge  A_2)}}
\newcommand{\wffr}{\ensuremath{(A_1\wedge \neg A_2)}}
\newcommand{\wffltwo}{\ensuremath{(\neg A_3\wedge  A_4)}}
\newcommand{\wffrtwo}{\ensuremath{(\neg A_3\wedge \neg A_4)}} 
\newcommand{\Aone}{\ensuremath{A_1}}
\newcommand{\nAone}{\ensuremath{\neg A_1}}
\newcommand{\Atwo}{\ensuremath{A_2}}
\newcommand{\nAtwo}{\ensuremath{\neg A_2}}  
\newcommand{\Athree}{\ensuremath{A_3}}
\newcommand{\nAthree}{\ensuremath{\neg A_3}}
\newcommand{\Afour}{\ensuremath{A_4}}
\newcommand{\nAfour}{\ensuremath{\neg A_4}} 
\newcommand{\singlemu}{\set{A_1,\neg A_3}}
\newcommand{\allmus}{
  \set{
    \set{A_1,A_2,\neg A_3,A_4},
    \set{A_1,A_2,\neg A_3,\neg A_4},
    \set{A_1,\neg A_2,\neg A_3,A_4},
    \set{A_1,\neg A_2,\neg A_3,\neg A_4}}}
\begin{figure}[t]  
  \centering 
\scalebox{.8}{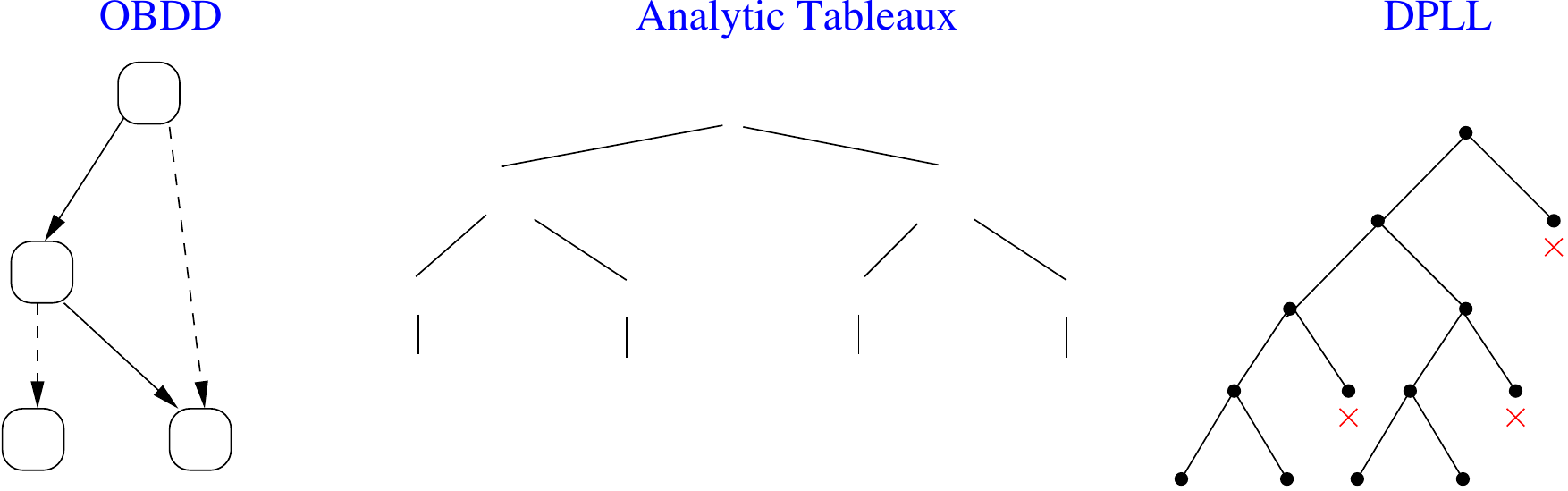} 
\caption{Left: OBDD for $\vi\defas\mainwff{}$. \newline
Center and right: Assignment enumeration for  $\vi$ via Analytic
Tableaux and (non-CNF) DPLL.
\label{fig:comparison}} 
\end{figure} 

\begin{example}
Consider $\vi\defas\mainwff$. 
  Figure~\ref{fig:comparison} 
represents the OBDD for $\vi$ and 
the search trees corresponding to 
All-SAT executions of Analytic
Tableaux and 
(non-CNF) DPLL on $\vi$. 
\begin{IGNOREINSHORT}
\footnote{Here in DPLL the pure-literal rule \cite{davis7} is not used because 
 in All-SAT it may hinder the enumeration of
 some relevant models (see, e.g., \cite{sebastiani_frocos07}).}  
\end{IGNOREINSHORT}
The former produces the single assignment
\singlemu{}. Notice that $\singlemu\entails\vi$ but $\singlemu\not
\validates\vi$.  The latter two produce the assignments \allmus{}.

Notice that neither Analytic Tableaux nor DPLL can produce \singlemu{} alone.
\hfill $\diamond$
\end{example}

One may argue that in SAT/AllSAT the distinction between $\validates$ and
$\entails$ is not much relevant in practice, because we can
CNF-ize upfront the input formulas ---typically by
variants of Tseitin CNF-ization--- removing tautological clauses, and this
distinction does not hold for (tautology-free) CNF
formulas.
However, we notice that with Tseitin CNF-ization we may loose
information regarding entailment or \verbvalidation{}. In fact, if
$\psi$ on $\alla\cup\allb$ is the result of Tseitin CNF-izing $\vi$,
then:
\begin{itemize}
\item 
$\mu\validates\vi$ \emph{does not imply} that there exists a total assignment
  $\delta$ on \allb s.t. $\mu\cup\delta\validates\psi$;
\item 
$\mu\entails\vi$ \emph{does not imply} that there exists a total assignment
  $\delta$ on \allb s.t. $\mu\cup\delta\entails\psi$.
\end{itemize}

\begin{example}
Consider $\vi\defas A_1 \vee (A_2\wedge A_3)$ 
and its Tseitin CNF-ized version:\\
$
\begin{array}{lll}
\psi\defas & (A_1\vee  B_1)\ \wedge \\
& (\neg B_1 \vee  A_2) \wedge
(\neg B_1 \vee  A_3) \wedge 
(B_1\vee \neg A_2 \vee \neg A_3) & \ \ \ //\ B_1 \iff (A_2\wedge A_3). \\
\end{array}
$\\
On the one hand, $\mu\defas \set{A_1}$ is such that $\mu\validates\vi$. 
On the other hand, there is no total truth assignment $\delta$ on
\set{B_1} s.t. $\mu\cup\delta\validates\psi$. In fact, neither 
$\set{A_1,B_1}\validates\psi$ nor $\set{A_1,\neg B_1}\validates\psi$.

Consider $\vi\defas
(A_1\wedge \pos A_2)\vee (A_1\wedge\neg A_2) 
$ 
and its Tseitin CNF-ized version:\\
$
\begin{array}{lll}
\psi\defas  
& (B_1\vee B_2)\ \wedge & \\
& (\neg B_1 \vee \pos A_1) \wedge
(\neg B_1 \vee \pos A_2) \wedge
(B_1\vee \neg A_1 \vee \neg A_2)\ \wedge\ & \ \ \ //\ B_1 \iff (A_1\wedge \pos A_2) \\
& (\neg B_2 \vee \pos A_1) \wedge
(\neg B_2 \vee \neg A_2) \wedge
(B_2\vee \neg A_1 \vee \pos A_2).\ & \ \ \ //\ B_2 \iff (A_1\wedge \neg A_2) 
\end{array}
%
$\\
Although $\mu\defas \set{A_1}$ is such that $\mu\entails\vi$,
 there is no total truth assignment $\delta$ on
\set{B_1,B_2} s.t. $\mu\cup\delta\entails\psi$:
$\set{A_1,B_1,\neg B_2}\not\entails \psi$ since
$\set{A_1,\neg A_2,B_1,\neg B_2}\not\entails \psi$;
$\set{A_1,\neg B_1,B_2}\not\entails \psi$
since $\set{A_1,A_2,\neg B_1,B_2}\not\entails \psi$;
$\set{A_1,B_1,B_2}$ and $\set{A_1,\neg B_1,\neg B_2}$ are both inconsistent
with $\psi$.
\hfill $\diamond$
\end{example}

\ignore{
\begin{proposition}
 \RSTODO{PROVA e scrivi controesempi. Scrivi in background l'equivalente controparte per $\eta \models\vi$\\}
Let $\vi$ be a generic (non-CNF)
formula on \alla and let $\psi$ on $\alla\cup\allb$ be the result of
CNF-izing \vi by applying standard Tseitin CNF-ization 
\cite{tseitin1} and removing valid clauses from $\psi$ if any, \allb
being the set of fresh propositions introduced in the process. Then:
\begin{itemize}
\item 
If $\mu\cup\gamma\validates\psi$ for some total truth assignment
  $\gamma$, then $\mu\validates\vi$.
\item 
If $\mu\cup\gamma\entails\psi$ for some total truth assignment
  $\gamma$, then $\mu\entails\vi$.
\item 
$\mu\validates\vi$ \emph{does not imply} that there exists a total assignment
  $\delta$ on \allb s.t. $\mu\cup\delta\validates\vi$
\item 
$\mu\entails\vi$ \emph{does not imply} that there exists a total assignment
  $\delta$ on \allb s.t. $\mu\cup\delta\entails\vi$
\end{itemize}

\end{proposition}
}

\section{Dealing with existentially-quantified formulas}
\label{sec:partialsat-exist}
In many application domains, fundamental operations ---like {\em 
  preimage computation} in symbolic model checking
(see e.g. \cite{burch1}) or {\em predicate abstraction} in SW verification
(see e.g. \cite{graf_predabs97,beyercgks09})---
require dealing with existentially-quantified formulas and with
the enumeration of partial assignments satisfying them.
Thus,
we lift the discussion of \sref{sec:partialsat} to
existentially-quantified formulas, and
we wish to provide a satisfactory definition of partial-assignment
satisfiability for an existentially-quantified propositional formula.
\RSTODO{non lo facciamo pero'... cambiare la narrazione.}

One first possibility is to see partial-assignment satisfiability as
{\em \verbvalidation}, leveraging Definition~\ref{def:validation}
and Property~\ref{prop:total-shannon} to the existentially-quantified
case. By \eqref{eq:shannon} and
Definition~\ref{def:validation},
$\mu\validates{} \shannonpsib$ iff
\apply{(\shannonpsib)}{\mu} is $\top$, that is, iff there exists some
 $\delta_i$ s.t. \apply{\apply{\psi}{\delta_i}}{\mu} is $\top$, that
 is, iff there exists some $\delta_i$ s.t. $\mu\cup\delta_i$ \verbvalidates{} $\psi$. 
This leads to the following definition and relative property. 
\begin{definition}
\label{def:validation-exist}
 We say that a {\em partial} truth assignment $\mu$ on $\alla$ {\bf
   \verbvalidates} $\exists
\allb.\psi$ if and only if, there exists a total truth assignment 
$\delta$ on $\allb$ s.t.
\ignore{the result of applying $\mu\cup\delta$ to 
$\psi$ is $\top$, that is,}
$\mu\cup\delta\validates\psi$.
\end{definition}

\begin{property}
Let   $\psi$ be a formula on
$\alla\cup\allb$ and $\mu$ be a partial assignment on \alla.
Then\\
  $\mu\validates\exists \allb. \psi$ iff $\mu\validates\shannonpsib$. 
\end{property}

One second possibility is to see partial-assignment satisfiability as
{\em entailment}, leveraging Definition~\ref{def:entailment}
and Property~\ref{prop:total-shannon} to the existentially-quantified
case. We notice that $\mu\entails \shannonpsib$ iff, for every total
assignment $\eta$ s.t. $\eta\supseteq\mu$, $\eta\models\shannonpsib$,
that is, by Property~\ref{prop:total-shannon} for every total
assignments $\eta$ s.t. $\eta\supseteq\mu$ exists a total assignment 
$\delta$ on \allb s.t. $\eta\cup\delta\models\psi$.
This leads to the following definition and relative property. 
\begin{definition}
\label{def:entailment-exist}
We say that a {\em partial} truth assignment $\mu$ on $\alla$ {\bf entails}
$\exists \allb.\psi$, written $\mu\entails\exists \allb.\psi$,
if and only if , for every total truth assignment $\eta$ on $\alla$ 
extending $\mu$,
there exists a total truth assignment $\delta$ on $\allb$
s.t. $\eta\cup\delta$ satisfies $\psi$. 
\end{definition}

\begin{property}
Let   $\psi$ be a formula on
$\alla\cup\allb$ and $\mu$ be a partial assignment on \alla.
Then\\
  $\mu\entails\exists \allb. \psi$ iff $\mu\entails\shannonpsib$. 
\end{property}

  \noindent
  Notice the nesting order of forall/exists in
Definition~\ref{def:entailment-exist}: ``for every $\eta$ exists $\delta$
s.t. ...''. In fact, distinct $\eta$'s may satisfy distinct disjuncts 
$\apply{\psi}{\delta_i}$ in \shannonpsib{}, requiring thus distinct 
$\delta_i$'s.
\ignore{Therefore, we notice that the following statement, inverting the
quantification order, is \emph{not} a suitable definition for $\mu\entails\exists \allb.\psi$:
%
\\ \noindent
  ``{\em We say that a {\em partial} truth assignment $\mu$ on $\alla$ entails
  $\exists \allb.\psi$ if and only if, there exists a total truth
  assignment $\delta$ on $\allb$ s.t., for every total truth assignment
  $\eta$ extending $\mu$, $\eta\cup\delta$ satisfies $\psi$.}''
}
%

Due to Proposition~\ref{prop:stronger} and
Property~\ref{prop:total-shannon} we have that
$\mu\validates\exists\allb.\psi$ 
 (Definition~\ref{def:validation-exist}) is strictly
stronger than $\mu\entails\exists\allb.\psi$ 
 (Definition~\ref{def:entailment-exist}).
Remarkably, and unlike with the un-quantified case, this is the case 
 \emph{even if $\psi$ is a tautology-free CNF formula}!
(Intuitively, this can be seen as a consequence of the fact that 
\shannonpsib is not in CNF even if $\psi$ is in CNF.)

\begin{example} Consider $\mu\defas\set{A_1}$
  and the tautology-free CNF formula on $\alla\cup\allb$:\\
$
\begin{array}{lll}
\psi\defas  
& (B_1\vee B_2)\ \wedge & \\
& (\neg B_1 \vee \pos A_1) \wedge
(\neg B_1 \vee \pos A_2) \wedge
(B_1\vee \neg A_1 \vee \neg A_2)\ \wedge\ & \ \ \ \\
& (\neg B_2 \vee \pos A_1) \wedge
(\neg B_2 \vee \neg A_2) \wedge
(B_2\vee \neg A_1 \vee \pos A_2).\ & \ \ \ 
\end{array}
$\\
Then we have that 
$\shannonpsib=
(A_1\wedge A_2\wedge\neg A_2)\vee (A_1\wedge A_2)\vee (A_1\wedge\neg A_2) 
\vee \bot
$ 
so that $\mu\entails\shannonpsib$ but 
$\mu\not\validates\shannonpsib$.
Thus, we have that $\mu\entails\exists\allb.\psi$ but
$\mu\not\validates\exists\allb.\psi$.
\hfill $\diamond$
\end{example}


\subsection{A Relevant Example Application: Predicate Abstraction.}

 
Given a propositional formula $\phi$ on $\allb$ and a set
$\mathbf{\Phi}\defas\set{\phi_i}_{i}$ of formulas on $\allb$ denoting relevant
``predicates'' and a set $\alla$ of fresh proposition s.t. each $A_i$
labels $\phi_i$, 
then the {\em Predicate Abstraction} of $\phi$ wrt. $\mathbf{\Phi}$ is
defined as follows \cite{graf_predabs97}:
\begin{eqnarray}
  \label{eq:predabs}
 \predabsphi \defas
\exists \allb. 
( \phi \wedge \bigwedge_{i} (A_i\iff \phi_i)
) .
\end{eqnarray}
 \predabsphi{}
is typically computed as  disjunction of mutually-inconsistent partial
assignments (cubes) $\mu_j$ on \alla s.t. $\mu_j\entails
\predabsphi$ and 
$\bigvee_j\mu_j$ is equivalent to \predabsphi
\cite{allsmt,cavada_fmcad07_predabs}.
\footnote{Notice that predicate Abstraction is most often referred to SMT formulas
  $\phi$ and $\set{\phi_i}_i$, so that \eqref{eq:predabs} involves
  also the existential quantification of first-order theory-specific variables
  and $\mu_i$ are theory-consistent SMT assignments
  \cite{graf_predabs97,allsmt,cavada_fmcad07_predabs}.
However, restricting our discussion to the purely-propositional case suffices for our
purposes and makes the explanation much simpler.}

We notice that in the computation of such cubes the distinction between $\validates$ and
$\entails$ may be very relevant: whereas 
it would be desirable to look for partial assignments $\mu_j$ 
{\em entailing} \predabsphi{}  to keep them small and hence reduce
their number,
 most algorithms can reveal
only when $\mu_j$ 
{\em \verbvalidates{}} it, and 
are thus
incapable of producing partial assignments 
$\mu_j$ s.t. $\mu_j\entails\predabsphi$ and
$\mu_j\not\validates\predabsphi$.
This happens every time that, 
for some $A_k$ and some $\mu^k$
on (subsets  of) $\alla\setminus \set{A_k}$,
{\em both} 
$\apply{(\phi\wedge\bigwedge_{i\neq k}(A_i\iff \phi_i))}{\mu^k} \wedge \phi_k$
and 
$\apply{(\phi\wedge\bigwedge_{i\neq k}(A_i\iff \phi_i))}{\mu^k} \wedge
\neg \phi_k$
 are satisfiable but they are satisfied by {\em distinct} sets of assignmets
 $\delta$ on \allb (Definition~\ref{def:entailment-exist}), so that
 $\mu_k\entails\predabsphi$ but 
 $\mu_k\not\validates\predabsphi$.

\begin{example}
Consider the CNF formula
$\phi\defas(\neg B_1\vee B_2)\wedge (B_1\vee \neg B_2)$
and the ''predicate'' CNF formulas
$\Phi_1\defas B_1\wedge B_2$ and
$\Phi_2\defas \neg B_1\wedge B_2$.
Then
\begin{eqnarray}
  \label{eq:predabs1}
  \predabsphi &\defas & \exists B_1 B_2. \left (
 \begin{array}{lll}
(\neg B_1\vee B_2)\wedge (B_1\vee \neg B_2) & \wedge \\
(A_1 \iff (B_1\wedge B_2)) & \wedge \\
(A_2 \iff (B_1\wedge \neg B_2))
 \end{array}
  \right )
  \\
  \label{eq:predabs2}
  &\Leftrightarrow& (A_1 \wedge \neg A_2) \vee (\neg A_1 \wedge \neg A_2) 
  \\
  \label{eq:predabs3}
  &\Leftrightarrow& \neg A_2
\end{eqnarray}
Both \set{A_1,\neg A_2} and \set{\neg A_1,\neg A_2}
\verbvalidate{} \predabsphi,
whereas  \set{A_1} entails it without \verbvalidating{} it.
Thus, if the algorithm is able to detect if
$\mu_j\entails\predabsphi$ and
$\mu_j\not\validates\predabsphi$, then it can return \eqref{eq:predabs3},
otherwise it can only return \eqref{eq:predabs2}.
\hfill $\diamond$
\end{example}

%
\noindent
Therefore, having algorithms able to stop extending $\mu_j$ as soon as 
$\mu_j\entails\predabsphi$,  even when 
$\mu_j\not\validates\predabsphi$,  would produce much more compact
 formulas.

\ignore{
\section{A simple poll}
\label{sec:poll}

\pagestyle{plain}
\pagenumbering{roman}

\pagestyle{plain}
\pagenumbering{arabic}

\title{%
  A Poll on the Meaning of Partial-Assignment Satisfiability
%
%
}

\author{
Roberto Sebastiani 
}

\institute{%
DISI, University of Trento, Italy%
}

\date{\today}

\maketitle
\ignoreinshort{
\large
\begin{center}
\noi
{\em Latest update: \today}
\end{center}
}

\begin{center}
\today
\end{center}


\label{sec:}

\section*{Premise}
This document reports the text and results of a poll I made among 65
members of current and past PC of SAT and SMT conference on the notion
of ``partial-assignment satisfiability''. Unfortunately only 20 people
replied. I asked each person to say
whether the following statements  were true or false according his/her
intended meaning of the sentence  ``a partial truth assignment
satisfies...''.

\section*{Statements}

\begin{enumerate}

\item
A {\em partial} truth assignment $\mu$ on $\alla$ satisfies $\vi$
if and only if all the total truth assignments $\eta$ on $\alla$ 
which extend $\mu$ satisfy $\vi$.

\item
A {\em partial} truth assignment $\mu$ on $\alla$ satisfies $\vi$
if and only if the result of applying $\mu$ to $\vi$ is $\top$.

\item
Let $\vi$ and $\vione$ be logically equivalent. 
Then a {\em partial} truth assignment $\mu$ on $\alla$ satisfies $\vi$
if and only if $\mu$  satisfies $\vione$.


\ignore{
\item
Checking
if a {\em partial} truth assignment $\mu$ on $\alla$ satisfies $\vi$
requires at most a polynomial amounts of steps.
}

\item
A {\em partial} truth assignment $\mu$ on $\alla$ satisfies $\exists \allb.\ps$
if and only if, there exists a total truth assignment $\gamma$ on $\allb$
s.t., for every total truth assignment $\eta$ extending $\mu$,
$\eta\cup\gamma$ satisfies $\ps$. 

\item
A {\em partial} truth assignment $\mu$ on $\alla$ satisfies $\exists
\allb.\ps$ if and only if, there exists a total truth assignment 
$\gamma$ on $\allb$ s.t. the result of applying $\mu\cup\gamma$ to 
$\ps$ is $\top$.

\end{enumerate}

\section*{Legenda}
In the above statements:
\begin{itemize}
\item 
$\alla\defas\set{A_1,...,A_N}$,
$\allb\defas\set{B_1,...,B_K}$ 
are disjoint sets of propositional atoms,
\item  
$\vi$, $\vione$  are generic propositional formulas over $\alla$, 
\item 
 $\ps$ is a generic propositional formula over $\alla\cup\allb$, 
\item 
$\mu$ 
is a {\em partial} truth assignment on $\alla$ (i.e., $\mu$ assigns a 
value only to a subset of $\alla$)
\item 
$\eta$ 
is a {\em total} truth assignment on $\alla$ (i.e., $\eta$ assigns a 
value to all atoms in $\alla$)
\item 
$\gamma$ 
is a {\em total} truth assignment on $\allb$ (i.e., $\gamma$ assigns a 
value to all atoms in $\allb$)
\end{itemize}
By ``apply a (total or partial) assignment $\mu$ to $\vi$'' we mean 
``substitute all instances of each assigned
$A_i$ in $\vi$ with the truth value in $\set{\top,\bot}$  assigned by
$\mu$ and apply the standard
propagation of truth values through the Boolean connectives
(i.e., $\neg\top \thus \bot$, 
$\neg\bot \thus \top$,
$\phi\wedge\top \thus \phi$, 
$\phi\wedge\bot \thus \bot$, and the corresponding ones for $\vee$, $\imp$, $\iff$).''

\newpage
\section*{Results}
Here follow the complete results (table \ref{tab:results} left) and
its summary, stating the number of people providing each combination
of answers (table \ref{tab:results_summary} right).  
All the 20 repliers have been pointed to this document.
The names of the repliers are kept anonymous. 
To each person who replied, however, I gave a code number so that he/she can
crosscheck his/her answer in the table. The code number is completely
uncorrelated with the alphabetic ordering of the names or surnames of
the repliers.

\begin{table}[t]
  \begin{tabular}{lll}
  \begin{tabular}{|r|l|l|l|l|l|}
\hline
Replier Code  &    \#1 & \#2 & \#3 &  \#4 & \#5\\
\hline
1 & true & false & true & true & true\\
2 & true & true & true & true & true\\
3 & false & true & false & false & true\\
4 & true & false & true & true & false\\
5 & true & false & true & true & false\\
6 & true &  & true & false & \\
7 & false & true & false & false & true\\
8 & false & true & false & false & true\\
9 & true & true & true & false & false\\
10 & false & true & false & false & true\\
11 & false & true & false & false & true\\
12 & true & true & false & true & true\\
13 & false & true & true & false & true\\
14 & true & false & true & false & false\\
15 & true & true & true & true & true\\
16 & true & true & true & false & false\\
17 & true & false & true & true & false\\
18 & true & false & true & true & false\\
19 & false & true & true & false & true\\
20 & true & true &  & true & true\\
\hline
  \end{tabular}
  \vspace{.5cm}
& & 
  \centering
  \begin{tabular}{|l|l|l|l|l|r|}
    \hline
    \#1 & \#2 & \#3 &  \#4 & \#5 & \# of repliers\\
    \hline
false & true & false & false & true & 5\\
false & true & true & false & true & 2\\
\hline
true & false & true & false & false & 1\\
true &  & true & false &  & 1\\
true & false & true & true & false & 4\\
true & false & true & true & true & 1\\
\hline
true & true & false & true & true & 1\\
true & true & false & true & true & 2\\
true & true & true & true & true & 2\\
 true & true &  & true & true & 1 \\
\hline
 &  &  &  &  & 20\\
\hline
  \end{tabular}
  \end{tabular}
  \caption{Left: Complete results of the poll. The code number of the left is the
    code given to each person who replied for cross-checking his/her
    reply.
  A blank is when a replier omitted giving an answer to the specific
  question.
\newline Right: Summary of the results.}
  \label{tab:results}
  \label{tab:results_summary}
\end{table}

\section*{Comments}

Some comments are in order. The questions refer to a short
paper I wrote on the
topic  \cite{sat19}.
%
Statements 1, 2, 3, 4, 5 refer, respectively,
to Definition 1 (entailment), Definition 2 (\verbvalidation),
Property 3(ii), Remark 1, and Definition 4 in \cite{sat19}.

\begin{itemize}
\item \#1=true means that one sees partial-assignment
  satisfiability as entailment (Definition 1 in \cite{sat19}). 
\item \#2=true means that one sees partial-assignment
  satisfiability as \verbvalidation (Definition 2 in \cite{sat19}). 
\item According to Proposition 1 in \cite{sat19}, {\em for generic formulas}
  \#1=true and
  \#2=true are not compatible, the latter being strictly stronger than
  the former. (They are equivalent only for
  tautology-free CNF formulas.)
\item \#3=true holds iff \#1=true (entailment) but not iff
\#2=true (\verbvalidation). See properties 3(ii) and 4(ii). 
\item If we define entailment \resp{\verbvalidation} of $\exists\allb.\psi$ as
  the entailment \resp{\verbvalidation} of its Shannon Expansion, 
  then we see that \#4=true is 
  incompatible with both entailment and \verbvalidation, whereas \#5=true 
means that one sees partial-assignment
  satisfiability as \verbvalidation. (Definition 3 and 4 and
  Properties 5 and 6 in \cite{sat19})
\end{itemize}
\noindent
To this extent, we would have answered one of the two 
alternative configurations:

\begin{center}
  \begin{tabular}{|l|l|l|l|l|c|}
    \hline
    \#1 & \#2 & \#3 &  \#4 & \#5 & \\
    \hline
true & false & true & false & false & \mbox{``satisfiability is entailment''} \\
false & true & false & false & true & \mbox{``satisfiability is \verbvalidation''}\\
    \hline
  \end{tabular}  
\end{center}

\noindent
Of course \#4 and \#5 depend on the assumption of defining entailment \resp{\verbvalidation} of $\exists\allb.\psi$ as
  the entailment \resp{\verbvalidation} of its Shannon Expansion. 
Also, the result may be effected by the fact that maybe we did not stress
enough the fact that ``generic formulas'' included non-CNF ones.
\ \\

Overall, although only 20 out of 65 people replied, 
the result 
reveals the lack of general agreement among the repliers on the notion of
partial-assignment satisfiability and on its consequences: e.g.,
the repliers split evenly into supporters 
of entailment, of \verbvalidation, and people considering them equivalent.

\FloatBarrier
\bibliographystyle{abbrv} 
\bibliography{rs_refs,rs_ownrefs,rs_specific,sathandbook}

}
\section{Conclusions and Future Work}
\label{sec:concl}
We have shown that, when dealing with non-CNF formulas or with
existentially-quantified formulas, we may have two distinct 
 notions of partial-assignment satisfiability, {\em entailment} and
 {\em \verbvalidation},
 with different
 properties, and that adopting one or the other may influence the
 effectiveness of assignment-enumeration procedures.

In the next future we wish to investigate the adoption of
partial-assignment reduction techniques exploiting entailment rather
than \verbvalidation, in particular in AllSMT and predicate
abstraction.
A possible candidate can be Dualization \cite{DBLP:conf/ictai/MohleB18}.
Also, we wish to investigate the tradeoff between the cost of detecting
entailment wrt. the 
effectiveness in reducing the number of enumerated assignments, in
particular when the latter may have computationally-hard consequences (e.g. WMI
\cite{MorPasSeb17,morettin_aij19}).

\section*{Acknowledgements} 

The analysis described in this paper 
strongly benefitted from interesting discussions, either personal or via
email, with
Armin Biere,
Alessandro Cimatti,
Allen van Gelder,
David Mitchell,
Sibylle M\"ohle,
Laurent Simon,
Armando Tacchella, and
Stefano Tonetta,
whom are all warmly thanked.

\pagenumbering{roman}
\FloatBarrier
\bibliographystyle{abbrv} 
\bibliography{rs_refs,rs_ownrefs,rs_specific,sathandbook}


\end{document}